\documentclass[twocolumn,showpacs,preprintnumbers,amsmath,amssymb,superscriptaddress,floatfix,dcolumn]{revtex4}

\usepackage{graphicx}
\usepackage{dcolumn}
\usepackage{bm}

\begin{document}
\newcommand{\bra}[1]{\langle #1 |}
\newcommand{\ket}[1]{| #1 \rangle}
\newcommand{\vect}[1]{\mathbf #1}
\newcommand{\dAAr}{$d_{\alpha\mbox{-}^{36}{\rm Ar}}$}
\newcommand{\dCSi}{$d_{^{12}{\rm C}\mbox{-}^{28}{\rm Si}}$}
\newcommand{\ovlp}[2]{\langle #1 | #2 \rangle}
\newcommand{\parag}[1]{\begin{flushleft}{\it #1}\end{flushleft}}
\newcommand{\AAr}{$\alpha$-$^{36}$Ar}
\newcommand{\CSi}{$^{12}$C-$^{28}$Si}


\title{Clustering and Triaxial Deformations of $^{40}$Ca}

\author{Yasutaka Taniguchi}
\affiliation{Department of Physics, Kyoto University, Kyoto 606-8502, Japan}
\affiliation{Yukawa Institute for Theoretical Physics, Kyoto University,
Kyoto 606-8502, Japan} 
\author{Masaaki Kimura}
\affiliation{Institute of Physics, University of Tsukuba, Tsukuba 305-8571, Japan}
\author{Yoshiko Kanada-En'yo}
\affiliation{Yukawa Institute for Theoretical Physics, Kyoto University,
Kyoto 606-8502, Japan} 
\author{Hisashi Horiuchi}
\affiliation{Research Center of Nuclear Physics, Osaka University,
Ibaraki 567-0047, Japan} 

\date{\today}

\begin{abstract}
 We have studied the positive-parity states of $^{40}$Ca using
 antisymmetrized molecular dynamics (AMD) and the generator coordinate method
 (GCM). Imposing two different kinds of constraints on the variational
 calculation, we have found various kinds of $^{40}{\rm
 Ca}$ structures such as a deformed-shell structure, as well as $\alpha$-$^{36}$Ar and
 $^{12}$C-$^{28}$Si cluster structures. After the GCM calculation,
 we obtained a normal-deformed band and a superdeformed band
 together with their side bands associated with triaxial
 deformation. The calculated $B(E2)$ values agreed well with
 empirical  data. It was also found that the normal-deformed and
 superdeformed bands  have a non-negligible $\alpha$-$^{36}$Ar  cluster
 component and  $^{12}$C-$^{28}$Si cluster component,  respectively. 
 This leads to the presence of an  $\alpha$-$^{36}$Ar higher-nodal
 band occurring above  the  normal-deformed band.
\end{abstract}


\maketitle

\section{Introduction}
Nuclear dynamics possess various aspects depending on mass regions,
excitation energies, and so on. In light-weight nuclei, it is known that
clustering plays a significant role in the features of ground and excited
states \cite{hor72,fuj80}. On the other hand, in heavier nuclei, the
clustering effects are not clear, though many
theoretical \cite{mic98,sak98} and experimental \cite{yam98} studies have
been conducted.  In the $fp$-shell region, the focus should be on
proton-rich $N\sim Z$ nuclei, because such nuclei can have a clustered
structure comprising stable nucleus.  Moreover, proton-rich
nuclei have a large radius and stronger Coulomb repulsion,
and may also derive a cluster structure.  In the scope of this
research, the features of cluster structures in $^{40}$Ca are a key issue, because they are the
heaviest $N=Z$ stable nuclei, and many experimental data exist for this
nuclei.  In this paper, we have studied the structure of $^{40}$Ca
as a starting point to understand the structures of medium- and heavy-weight
$N\sim Z$ nuclei.  $^{40}$Ca has a typical double closed-shell structure
nucleus and has a spherical ground state.  However, it is known
that many kinds of deformed band appear in low energy regions.  The
first $K^\pi = 0^+$ band built on the $J^\pi = 0_2^+$ state (3.35 MeV) is
considered to be a normal-deformed (ND) state and the dominant configuration is
$4p$-$4h$  \cite{ger67}. The $K^\pi = 2^+$ band built on
$J^\pi = 2_2^+$ (5.25 MeV) exists just above the $K^\pi = 0^+$ band.  It
has been suggested that the ND band deforms triaxially and has the $K^\pi =
2^+$ side band due to triaxiality \cite{ger69,ger77}.  

The $\alpha$-$^{36}$Ar cluster structure has been studied for a long
time, because $^{40}$Ca is an analogue of $^{16}$O, which has
an $\alpha$-$^{12}$C cluster structure in the first $K^\pi = 0^+$ band, as
a double closed-shell nuclei. The local potential
model \cite{pal80,ohk88,rei90} and $\alpha$-$^{36}$Ar orthogonal
condition model (OCM) \cite{oga77,sak94} have been performed theoretically.  Ohkubo {\it et
al.} suggested that the first $K^\pi = 0^+$ band (ND) has an $\alpha$-$^{36}$Ar
structure, and predicted that its parity-doublet $K^\pi = 0^-$ band and
$\alpha$-$^{36}$Ar higher nodal band exist in highly excited
states \cite{ohk88}.  Sakuda {\it et al.} obtained the $K^\pi = 2^+$ band
as well as the $K^\pi = 0^+$ and $0^-$ states using the $\alpha$-$^{36}$Ar
OCM, and succeeded in reproducing $E2$ transition strengths \cite{sak94}.
Experimentally, the $\alpha$-$^{36}$Ar structure is studied through the
$^{36}$Ar($^6$Li,$d$)$^{40}$Ca reaction \cite{yam93,yam94}.  The states
in these $K^\pi = 0^+$ and $0^-$ bands are populated by the
$\alpha$-transfer reactions and have large $\alpha$ spectroscopic
factors \cite{yam93}.  In the experiments we describe, the $\alpha$-$^{36}$Ar
higher-nodal states were also observed \cite{yam94}.

It has been suggested that the states in the $K^\pi = 0^+$ rotational band built on the $J^\pi = 0_3^+$ state (5.21 MeV) have a $8p$-$8h$
configuration \cite{ger69}, and have been observed during experimental
work searching for the $8p$-$8h$ states with $^{32}$S($^{12}$C,
$\alpha$)$^{40}$Ca reactions \cite{mid72}.  Due to the strong population
in the multi-nucleon transfer data and the strong $E2$
transitions \cite{mac71}, the $0^+$ (5.21 MeV), $2^+$ (5.63 MeV) and
$4^+$ (6.54 MeV) bands have been thought to belong to the superdeformed (SD)
band with the dominant $8p$-$8h$ configuration.  Recently, by using
GAMMASPHERE array detectors, the level structure of the deformed bands
in $^{40}$Ca has been explored and many excited states up to high spin
have been discovered. This band was thus confirmed as the SD
band \cite{ide01}.  

Motivated by these the experimental observations, many theoretical microscopic
studies on deformed states of $^{40}$Ca have been performed
recently with the methods of Skyrme-Hartree-Fock (SHF) \cite{ina02}, 
SHF-BCS + GCM \cite{ben03}, spherical-basis AMD \cite{kan05} 
and the shell model \cite{cau07}.  Inakura
{\it et al.} performed cranked SHF calculations without
assuming axial symmetry, though energy levels were not
calculated \cite{ina02}.  Bender {\it et al.} performed SHF-BCS + GCM
calculations \cite{ben03}.  Although they calculated energy levels and
quadrupole transition strengths in the ND and SD bands, they could not study
triaxiality nor side bands because they assumed axial symmetry.  In
these studies, the relationship between deformed states and cluster
structure was not discussed.  It has been suggested that the SD state forms a $^{12}$C-$^{28}$Si-like cluster structure in spherical-basis AMD \cite{kan05}.
Within the spherical-basis AMD, triaxiality does not appear in ND
nor SD states.  

The purpose of the present study is to understand the clustering and
triaxial deformations in the low energy states of $^{40}$Ca in a unified manner. We use the
framework of AMD + GCM.  The basis functions of GCM are obtained by energy
variation after parity projection with constraints.  We adopted two
kinds of constraints.  One is a constraint on the quadrupole deformation
parameter $\beta$ ($\beta$-constraint) and the other is on the distance $d$
between clusters' centers of mass ($d$-constraint).  It has already been proven that the $d$-constraint is useful for obtaining various kinds
of clustering wave function, which are not computed within a simple
$\beta$-constraint \cite{tan04}.  For example, the $^8$Be($2\alpha$)-$^{12}$C
cluster structure in $^{20}$Ne is calculated with a $d$-constraint but not
with the $\beta$-constraint.  Also in the case of $^{40}$Ca, 
many kinds of cluster structure, for example $\alpha$-$^{36}$Ar,
$^8$Be($2\alpha$)-$^{32}$S and $^{12}$C-$^{28}$Si, can be calculated
with the $d$-constraint, although no
cluster structure is obtained in $^{40}$Ca with the $\beta$-constraint.  
We superposed mean-field-type and cluster-type wave functions calculated with $\beta$- and $d$-constraints respectively, and calculated energies and $E2$ transition strength.  We
analyzed the superposed wave functions in order to investigate clustering and triaxial deformations.

This paper is organized as follows. In the next section
(\S\ref{sec:formulation}), we explain the framework of this
study.  The calculated results and discussions are presented in
\S\ref{sec:results}, and lastly, we present a summary in \S\ref{sec:summary}. 

\section{Framework}
\label{sec:formulation}
\subsection{Wave Function and Hamiltonian}
We used the theoretical framework of AMD + GCM. In the present study, the AMD
wave function is a Slater determinant of triaxially deformed Gaussian
wave packets (deformed-basis AMD),  
\begin{subequations}
\label{AMD}
\begin{eqnarray}
\ket{\Phi_{\rm int}}& = &\hat{\cal A} \ket{\varphi_1,\  \varphi_2,\cdots,\varphi_A}, \\
\ket{\varphi_i}& = &\ket{\phi_i,\  \chi_i,\   \tau_i}, \\
\ovlp{\vect{r}}{\phi_i}& = &\prod_{\sigma = x, y, z} \left( \frac{2\nu_\sigma}{\pi} \right)^{\frac{1}{4}} \exp \left[ - \nu_\sigma \left( r_\sigma - \frac{Z_{i\sigma}}{\sqrt{\nu_\sigma}} \right)^2 \right], \nonumber\\
\\
\ket{\chi_i}& = &\alpha_i \ket{\uparrow} + \beta_i \ket{\downarrow},\\
\ket{\tau_i}&  = &\ket{p}\  {\rm or}\  \ket{n}. 
\end{eqnarray}
\end{subequations}
Here, the complex parameters $\vect{Z}_i$, which represent the centroids
of the Gaussian in phase space, take independent values for each single
particle wave function.  The width parameters $\nu_x$, $\nu_y$ and
$\nu_z$ are real parameters and take independent values for each of the
$x$-, $y$- and $z$-directions, but are common for all nucleons.  The
spin part $\ket{\chi_i}$ is parametrized by $\alpha_i$ and $\beta_i$ and the
isospin part $\ket{\tau_i}$ is fixed as $\ket{p}$ (proton) or $\ket{n}$ (neutron).  The
quantities $(\vect{Z}_i,\alpha_i,\beta_i,\nu_x,\nu_y,\nu_z)$ are
variational parameters and are optimized by energy variation as
explained in the next subsection.  

The trial wave function in the energy variation with constraints is
a parity-projected wave function,  
\begin{equation}
\ket{\Phi^\pi} = \frac{1+\pi \hat{P}_r}{2} \ket{\Phi_{\rm int}}, 
\end{equation}
where $\pi$ is parity 
and $\hat{P}_r$ is the parity operator.  In this study, we will
discuss positive parity states.  

The Hamiltonian is,
\begin{equation}
\hat{H} = \hat{K} + \hat{V}_{\rm N} + \hat{V}_{\rm C} - \hat{K}_{\rm G}, 
\end{equation}
where $\hat{K}$ and $\hat{K}_{\rm G}$ are the kinetic energy and the
energy of the center of mass motion respectively, and $\hat{V}_{\rm N}$
is the effective nucleon-nucleon interaction.  We have used Gogny D1S
force (D1S) and Skyrme SLy7 force (SLy7) in the present work.  The Coulomb force
$\hat{V}_{\rm C}$ is approximated by a sum of seven Gaussians.  

\subsection{Energy Variation, Angular Momentum Projection and the
  Generator Coordinate Method} 
We performed energy variation and optimized the variational parameters
included in the trial wave function (Eqs. (\ref{AMD})) to find the state
that minimizes the energy of the system $E^\pi$,  
\begin{equation}
E^\pi = \frac{\bra{\Phi^\pi}\hat{H}\ket{\Phi^\pi}}{\ovlp{\Phi^\pi}{\Phi^\pi}} + V_{\rm cnst}. 
\end{equation}
Here, we add the constraint potential $V_{\rm cnst}$ to the expectation
value of Hamiltonian $\hat{H}$ in order to obtain the minimum energy
state under the optional constraint condition.  In this study, we
employed two types of constraint, which are on the  quadrupole
deformation parameter $\beta$ ($\beta$-constraint) and the distance
between clusters' centers of mass, $d$ ($d$-constraint) by employing
the potential $V_{\rm cnst}$,  
\begin{equation}
V_{\rm cnst} = 
\left\{
\begin{array}{ll}
v_{\rm cnst}^\beta ( \beta - \tilde{\beta} )^2 & \mbox{for $\beta$-constraint,}\\
v_{\rm cnst}^d ( d_{{\rm C}_m\mbox{-}{\rm C}_n} - \tilde{d}_{\rm C_m\mbox{-}C_n})^2          & \mbox{for $d$-constraint.}
\end{array}
\right. . 
\end{equation}
Here $\beta$ is the matter quadrupole deformation parameter, which is
defined in Ref.~\onlinecite{dot97}, and $d_{{\rm C}_m\mbox{-}{\rm C}_n}$
is the distance between the clusters' centers of mass ${\rm C}_m$ and
${\rm C}_n$,  
\begin{equation}
d_{{\rm C}_m\mbox{-}{\rm C}_n} = \left| \vect{R}_m - \vect{R}_n \right|,
\end{equation}
\begin{equation}
R_{n\sigma} = \frac{1}{A_n} \sum_{i\in {\rm C}_n} \frac{{\rm Re}
 Z_{i\sigma}}{\sqrt{\nu_\sigma}},  
\end{equation}
where $A_n$ is the mass number of cluster ${\rm C}_n$ and the
expression $i\in {\rm C}_n$ means that the $i$th nucleon is contained in
cluster C$_n$.  It should be noted that the $\sigma\ (=x, y, z)$
component of the spatial center of the single-particle wave function
$\ket{\varphi_i}$ is $\frac{{\rm Re} Z_{i\sigma}}{\sqrt{\nu_\sigma}}$.
When sufficiently large values are chosen for $v_{\rm cnst}^\beta$ and $v_{\rm
cnst}^d$, the resultant values $\beta$ and $d_{{\rm C}_m\mbox{-}{\rm
C}_n}$ of energy variation become $\tilde{\beta}$, $\tilde{d}_{{\rm
C}_m\mbox{-}{\rm C}_n}$, respectively.  We constrained the \dAAr\ and \dCSi\
values for the $d$-constraint.  In each calculation of energy variation, we
constrained one of $\beta$, \dAAr\ and \dCSi.  A detailed explanation
regarding the $d$-constraint may be found in Ref.~\onlinecite{tan04}.  

The energy variation with the AMD wave function is carried out using the
frictional cooling method \cite{kan95}.  The time evolution equation for
the complex parameters $\vect{Z}_i, \alpha_i$ and $\beta_i$ is  
\begin{equation}
\frac{dX_i}{dt} = - \mu_X \frac{\partial E^\pi}{\partial X_i^*},\  (i=1, 2, \cdots, A), 
\end{equation}
where $X_i$ is $\vect{Z}_i, \alpha_i$ or $\beta_i$, and that for the
real parameters $\nu_x, \nu_y$ and $\nu_z$ is  
\begin{equation}
\frac{d\nu_\sigma}{dt} = - \mu_\nu \frac{\partial E^\pi}{\partial
 \nu_\sigma},\  (\sigma = x, y, z).  
\end{equation}
The quantities $\mu_X$ and $\mu_\nu$ are arbitrary positive real numbers. 
The energy of the system decreases as time progresses, and after a
sufficient number of time steps, we obtain the minimum energy state.  

After the constrained energy variation for $\ket{\Phi^\pi}$, we
superposed the optimized wave functions employing the quadrupole
deformation parameter $\beta$ and the distances between the centers of mass
among clusters $d_{\rm C_m\mbox{-}C_n}$ for ${\rm C_m\mbox{-}C_n}$
configurations as the generator coordinate,  
\begin{eqnarray}
\ket{\Phi^{J^\pi}_M} 
= 
\sum_K &\hat{P}_{MK}^{J^\pi}&
\left(
\sum_i f_{iK}^\beta \ket{\Phi^\beta_i} 
\right. 
\nonumber\\
&&
\left.
+ \sum_{i,{\rm C_m\mbox{-}C_n}} f_{iK}^{d_{\rm C_m\mbox{-}C_{n}}}
\ket{\Phi^{d_{\rm C_m\mbox{-}C_{n}}}_i} 
\right)
\end{eqnarray}
where $\hat{P}_{MK}^{J^\pi}$ is the parity and total angular momentum
projection (AMP) operator, and $\ket{\Phi^\beta_i}$ and
$\ket{\Phi^{d_{\rm C_m\mbox{-}C_n}}_i}$ are optimized wave functions
with $\beta$- and $d_{\rm C_m\mbox{-}C_n}$-constraints for the constrained
values $\tilde{\beta}^{(i)}$ and $\tilde{d}_{\rm C_m\mbox{-}C_n}^{(i)}$
respectively.  The integrals over the three Euler angles included in
$\hat{P}_{MK}^J$ are evaluated by numerical integration.  The
coefficients $f_{iK}^\beta$ and $f_{iK}^{d_{\rm C_m\mbox{-}C_{n}}}$ are
determined by the Hill-Wheeler equation,  
\begin{equation}
\delta \left( \bra{\Phi^{J^\pi}_M} \hat{H} \ket{\Phi^{J^\pi}_M} - \epsilon \ovlp{\Phi^{J^\pi}_M}{\Phi^{J^\pi}_M}\right) = 0. 
\end{equation}

\subsection{Squared Overlap}
\label{sec:SO}
We defined squared overlap $SO$ to estimate the overlap of a specific model
space in the superposed wave function.  Suppose the non-orthonormalized
wave functions $\ket{\Phi_i^X}$ span the functional space $\{ X\}$,
for example, $\{ X\} = \{\ket{\Phi^\beta_i}\}$,
$\{\ket{\Phi^{d_{\rm\alpha\mbox{-}^{36}Ar}}_i}\}$ or
$\{\ket{\Phi^{d_{\rm ^{12}C\mbox{-}^{28}Si}}_i}\}$.  Orthonormalized
wave functions $\ket{\tilde{\Phi}_\alpha^X}$ are obtained by performing
unitary transformations from $\ket{\Phi_i^X}$,  
\begin{subequations}
\begin{eqnarray}
&&\ket{\tilde{\Phi}^X_\alpha} = u_{\alpha i} \ket{\Phi_i^X}, \\ 
&&\ovlp{\tilde{\Phi}^X_\alpha}{\tilde{\Phi}^X_\beta} = \delta_{\alpha\beta}. 
\end{eqnarray}
\end{subequations}

Using the $\ket{\tilde{\Phi}^X_\alpha}$, squared overlap between
$\ket{\Phi}$ and the functional space $\{ X\}$, $SO$ is defined as  
\begin{equation}
SO = \sum_\alpha |\ovlp{\tilde{\Phi}^X_\alpha}{\Phi}|^2. 
\end{equation}

\section{Results and Discussions}
\label{sec:results}
\subsection{Various Structures Obtained with $\beta$- and $d$-constraints}
\begin{figure}[hbt] 
  \begin{center}
    \includegraphics[width=0.45\textwidth]{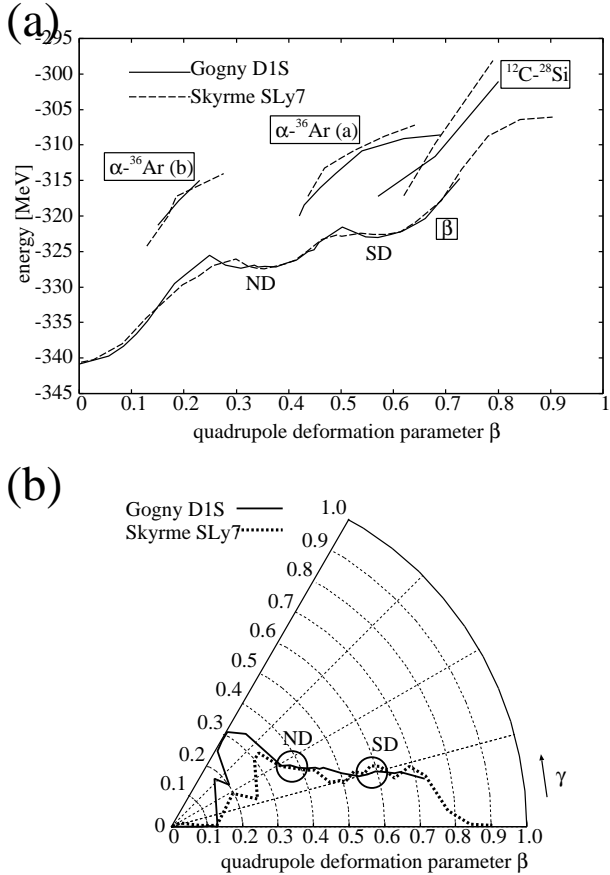}
    \caption{ (a): The energy curves for the positive-parity states
   obtained using $\beta$-, \dAAr- and \dCSi-constraints. The
   energies are plotted as functions of quadrupole deformation $\beta$.
   The solid (dotted) line shows the results for D1S (SLy7).\\
   (b): Projection of the energy curves onto the $\beta$-$\gamma$
   plane. The solid (dotted) line shows  the results for D1S (SLy7).}
    \label{fig:surface}
  \end{center}
\end{figure}
\begin{table}[bt]
  \begin{center} 
  \caption{Expectation values of kinetic and potential energies around the
   ground state, ND and SD minima. $K$, $V_{\rm NN}$, $V_{LS}$ and
   $V_{\rm C}$ denote kinetic energy, central, spin-orbit and Coulomb
   potentials.} 
  \label{tab:energy_components}
  \begin{tabular}{lccccccc}
    \hline\hline
interaction & $\beta$ & total & $K$ & $V_{\rm NN}$ & $V_{LS}$ & $V_{\rm C}$ \\
    \hline
Gogny D1S & 0.00 & $-340.9$ & 634.2 & $-1046.2 $ & $-0.1$ & 71.2 \\
Skyrme SLy7 & 0.00 & $-340.7$ & 637.9 & $-1050.0$ & $-0.0$ & 71.5\\
    \hline
Gogny D1S & 0.39 & $-326.7$ & 654.6 & $-1033.7$ & $-18.4$ & 70.9\\
Skyrme SLy7 & 0.39 & $-326.8$ & 653.7 & $-1031.9$ & $-19.4$ & 70.8\\
    \hline
Gogny D1S & 0.62 & $-322.2$ & 670.9 & $-1029.9$ & $-33.6$ & 70.4\\
Skyrme SLy7 & 0.61 & $-322.3$ & 676.2 & $-1035.9$ & $-33.3$ & 70.7\\
\hline\hline
  \end{tabular}
  \end{center}
\end{table}
We performed the energy variation after the projection to the
positive-parity state imposing two different kinds of constraints,
$\beta$- and $d$-constraints. 

Figure~\ref{fig:surface} (a) shows the obtained energy curves as
a function of matter quadrupole deformation $\beta$. By applying
the $\beta$-constraint we obtained energy curves for D1S (solid line)
and SLy7 (dotted line). Both forces give quite 
similar curves that have three local minima or shoulders at $\beta\sim
0, 0.4$ and 0.6. As shown in Table. \ref{tab:energy_components}, they
also give approximately the same kinetic and potential energies around each
minimum. Therefore, we mainly discuss the D1S result and make some
comments on the differences between D1S and SLy7 below. The lowest minimum at $\beta$=0 corresponds to the spherical
ground state and the two minima at $\beta\sim 0.4$ and 0.6 correspond to the
ND and SD states respectively. The excitation energies of the ND and SD minima
are approximately 14 MeV and 18 MeV. This result qualitatively agrees
with the constrained SHF calculation using SLy4
 \cite{ina02}. On the other hand, the energy
curve in the constrained SHF-BCS calculation
with  SLy6  \cite{ben03} reveals different behavior. It does not have the ND minimum nor shoulder, but has a SD
minimum.  The excitation energy of the SD minimum is much smaller than
ours and that of Ref. \onlinecite{ina02}.  This difference may be due to the strong pairing
correlation reported in Ref. \onlinecite{ben03}. 

\begin{figure}[hbt]
  \begin{tabular}{ccc}
    ND: $zy$-plane & ND: $zx$-plane & ND: $xy$-plane\\
    \begin{minipage}{0.15\textwidth}
      \begin{center}
	\includegraphics[width=\textwidth]{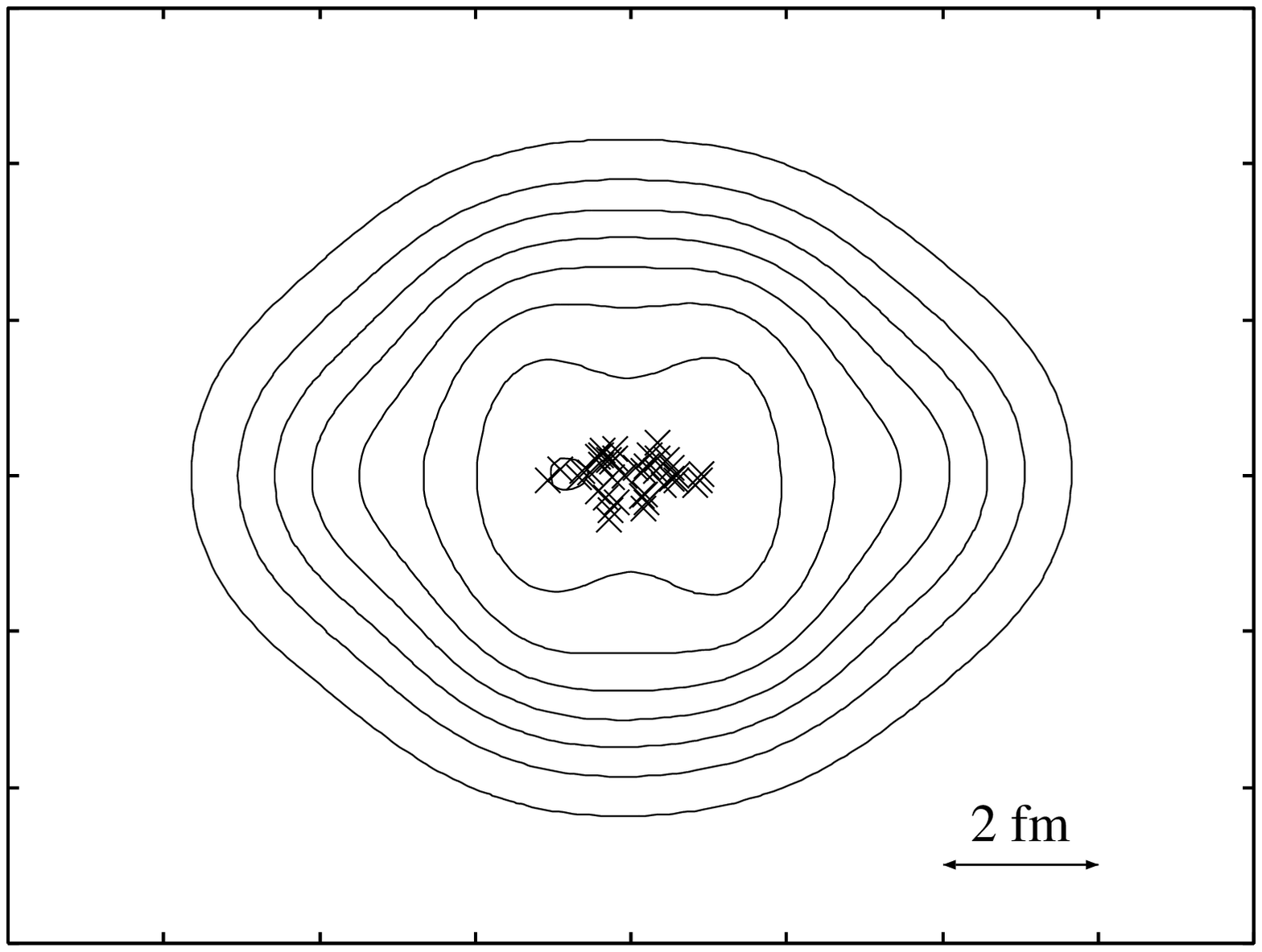}
      \end{center}
    \end{minipage}&
    \begin{minipage}{0.15\textwidth}
      \begin{center}
	\includegraphics[width=\textwidth]{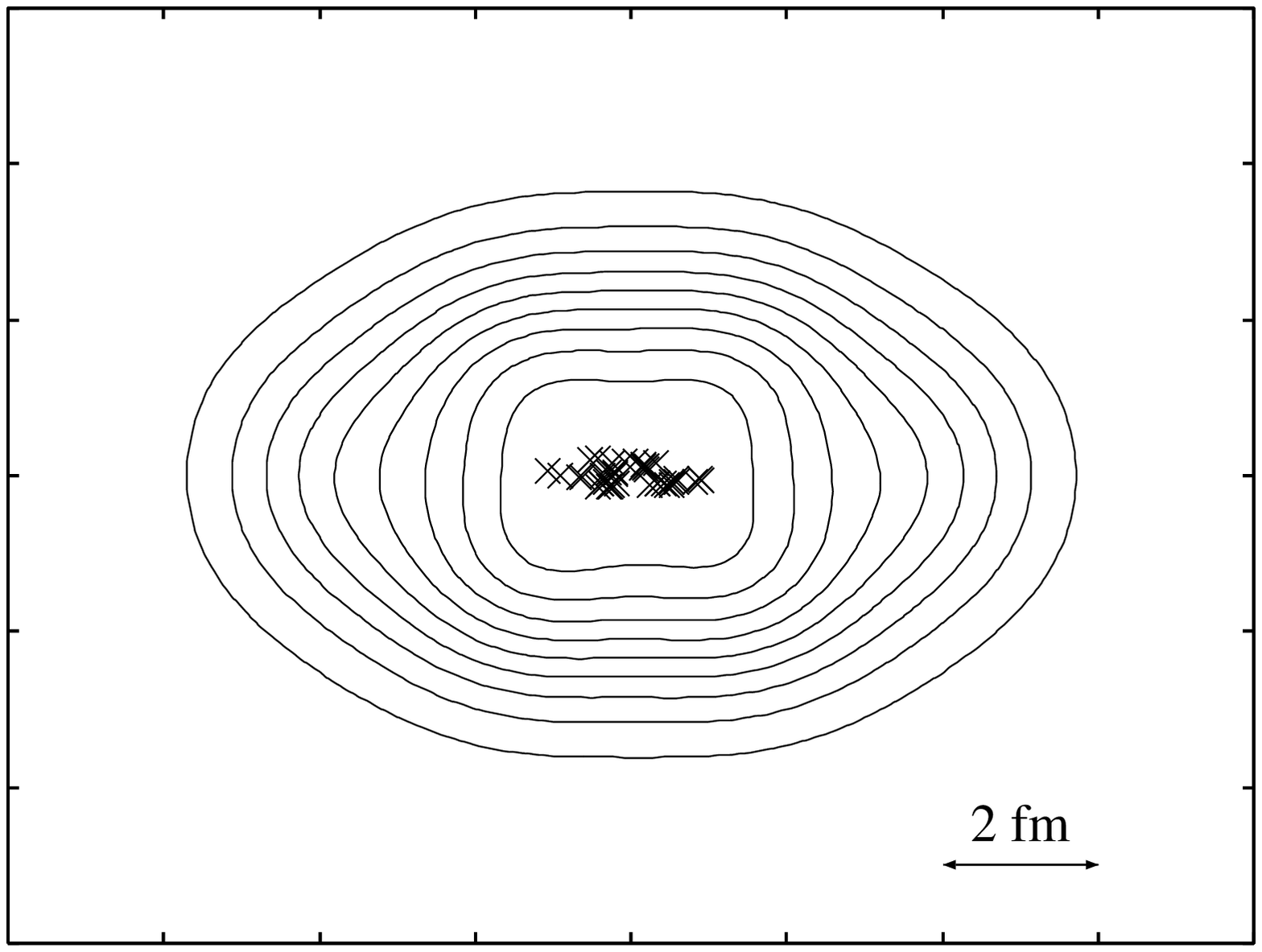}
      \end{center}
    \end{minipage}&
    \begin{minipage}{0.15\textwidth}
      \begin{center}
	\includegraphics[width=\textwidth]{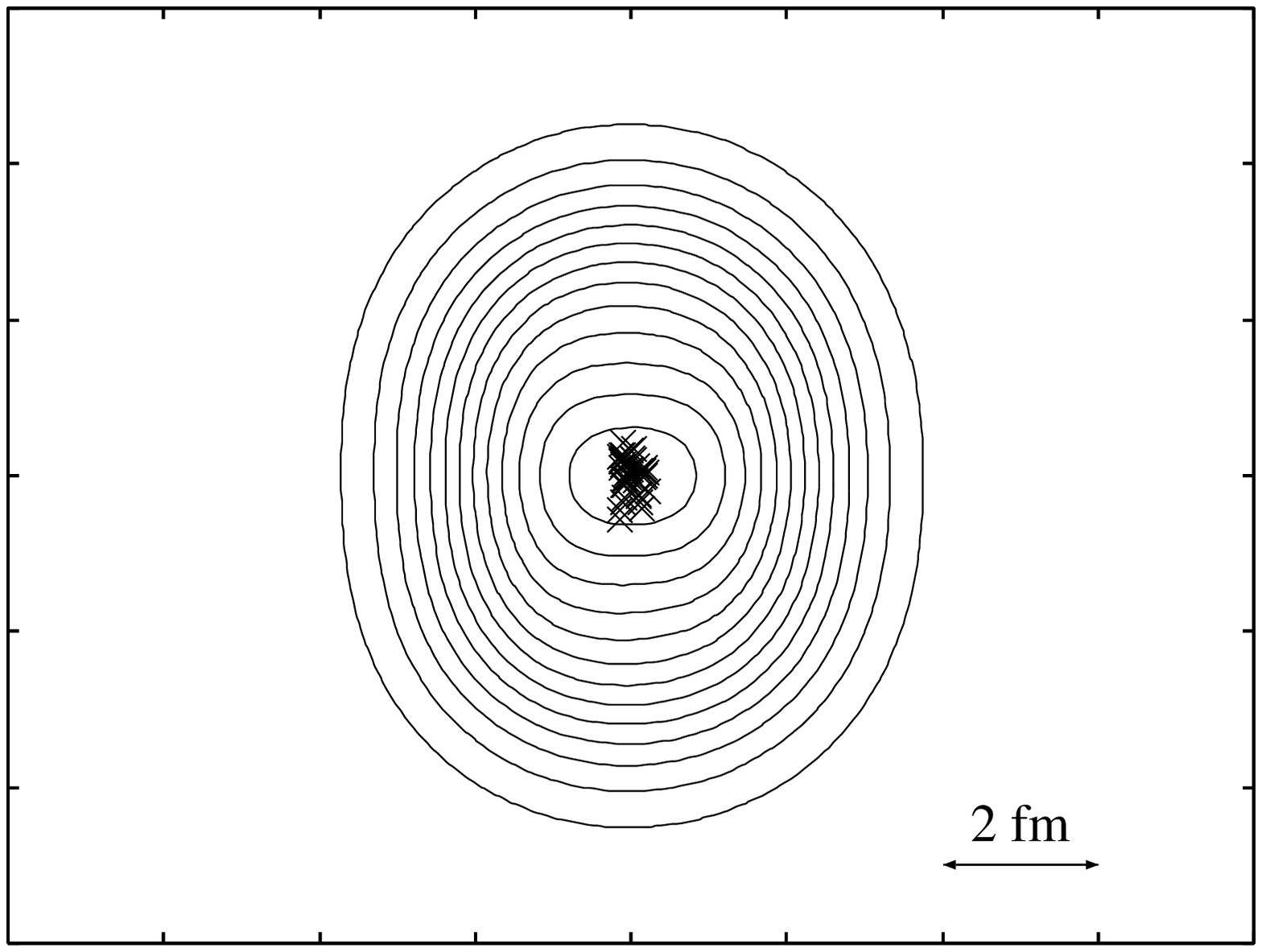}
      \end{center}
    \end{minipage}\\
    \\
    SD: $zy$-plane & SD: $zx$-plane & SD: $xy$-plane\\
    \begin{minipage}{0.15\textwidth}
      \begin{center}
	\includegraphics[width=\textwidth]{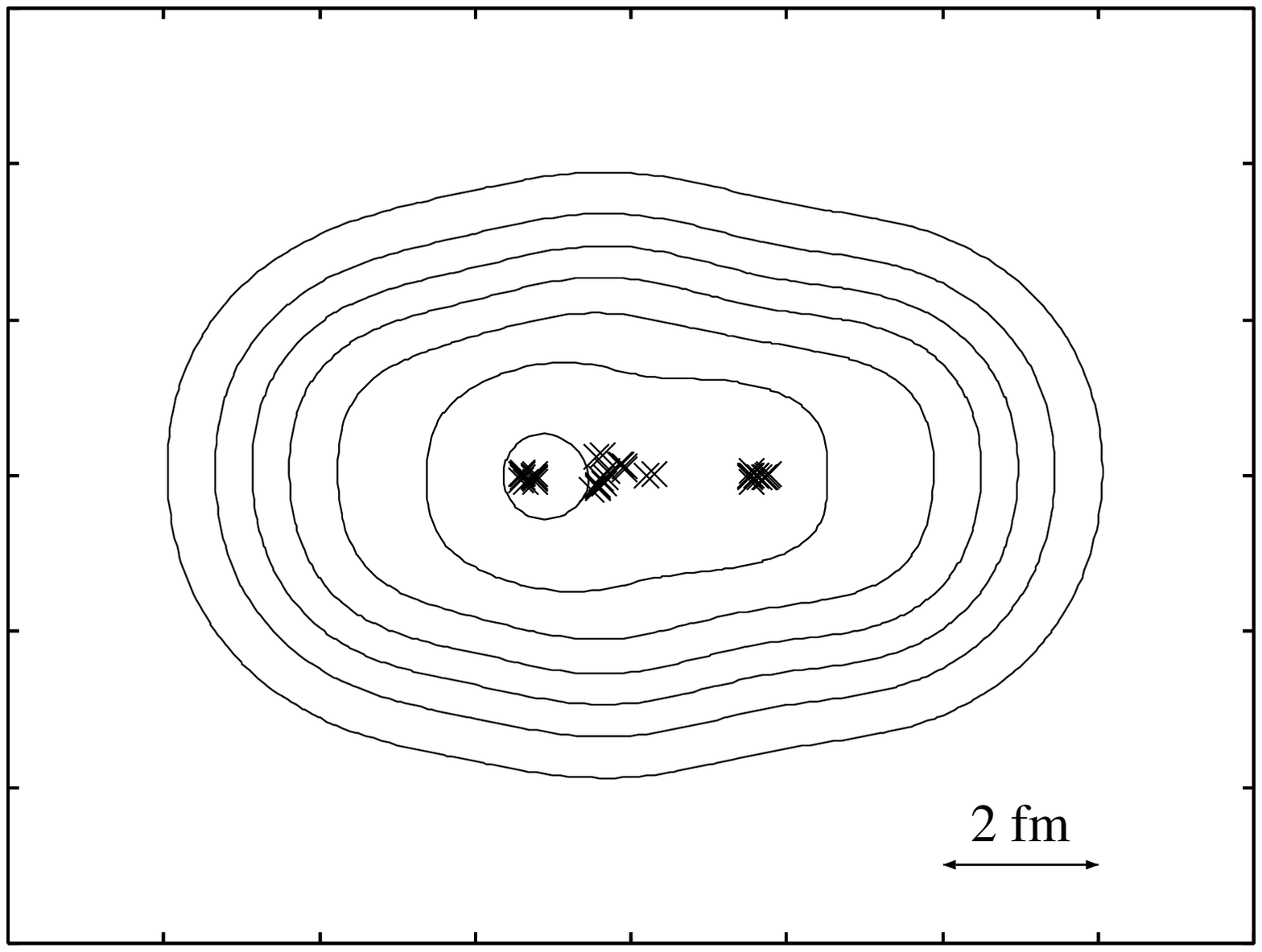}
      \end{center}
    \end{minipage}&
    \begin{minipage}{0.15\textwidth}
      \begin{center}
	\includegraphics[width=\textwidth]{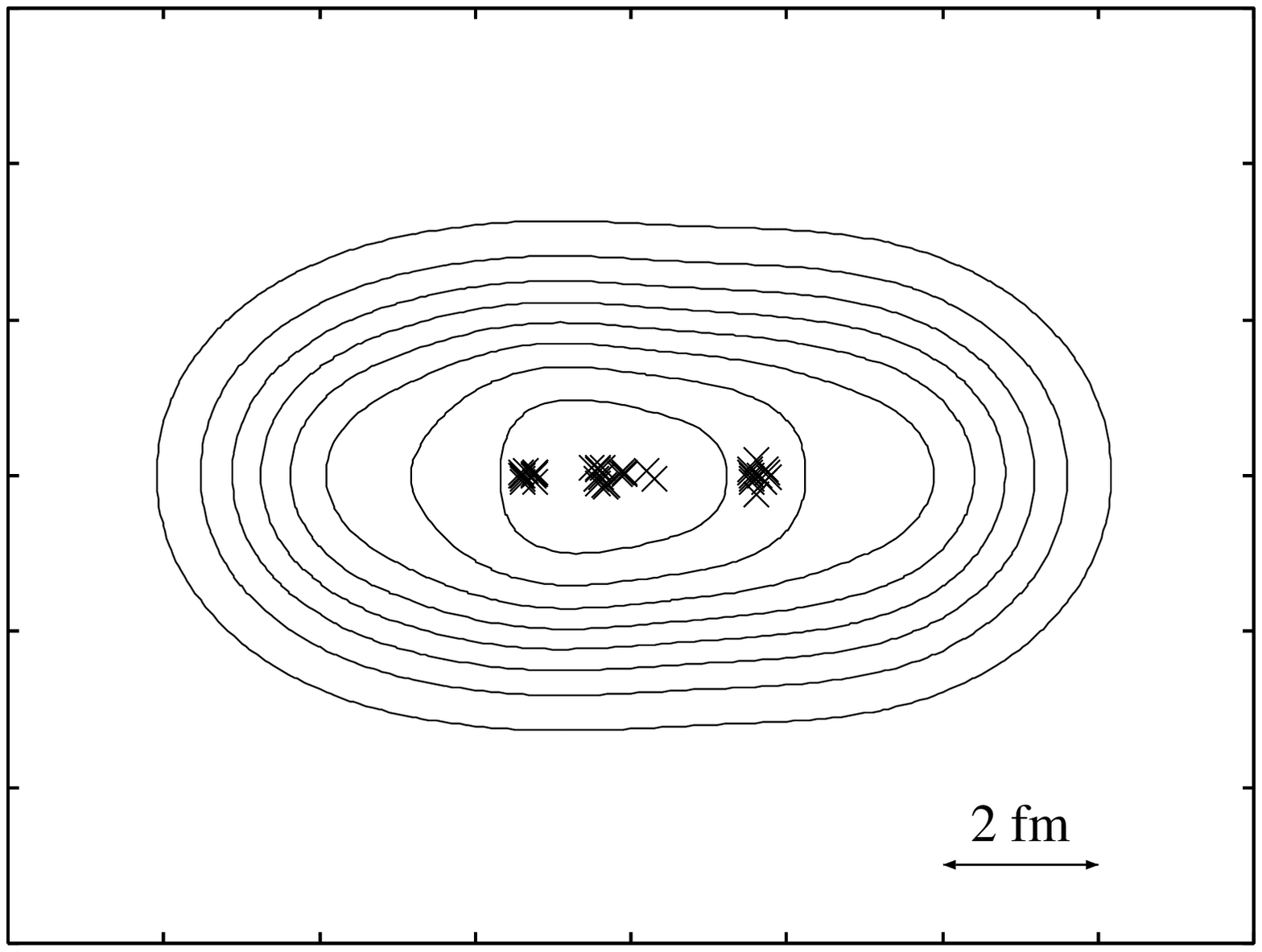}
      \end{center}
    \end{minipage}&
    \begin{minipage}{0.15\textwidth}
      \begin{center}
	\includegraphics[width=\textwidth]{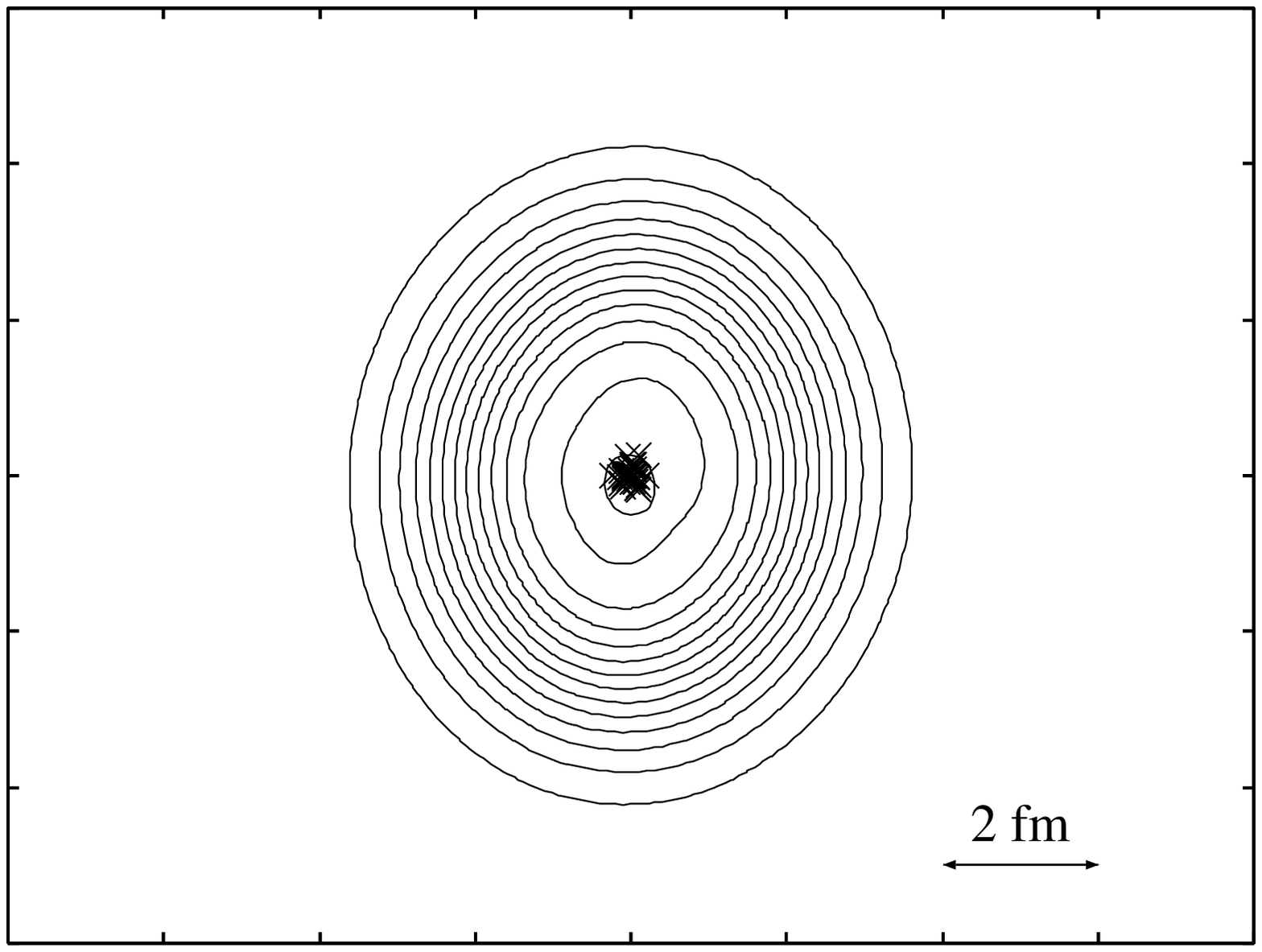}
      \end{center}
    \end{minipage}
  \end{tabular}
  \caption{Density distributions of the ND and SD minima obtained using the
 $\beta$-constraint. The crosses in the figure show the centroids of
 the wave packets. 
The deformation parameters are $(\beta, \gamma)=(0.39, 25.2^\circ)$ 
and $(0.62, 14.5^\circ)$ for the ND and SD minima, respectively. 
}
  \label{fig:density_beta}
\end{figure}

 In the present calculation, we do not make any assumptions regarding the
spatial symmetry of the wave function nor put any constraints on
the quadrupole deformation $\gamma$. $\gamma$ therefore has the optimal
value for each given value of $\beta$. Figure \ref{fig:surface}~(b)
shows the projection of the energy curve onto the $\beta$-$\gamma$
plane. It shows that in most regions the system is triaxially deformed
and the degree of the triaxial deformation greatly changes as a function
of $\beta$. Starting from the spherical ground state, the system rapidly
changes to oblate deformation around $\beta$=$0.15\mbox{-} 0.25$ via a small
prolate deformation. Then, it changes to triaxial deformation around
$\beta$=0.35 where the ND minimum appears and the system has the largest
triaxiality. With a further increase of $\beta$, $\gamma$ decreases
gradually from $\gamma$=$30^\circ$ at $\beta$=0.35 to
$\gamma$=$15^\circ$ at $\beta$=0.70. The SD minimum that appears at
$\beta$=0.60 also has a large triaxial deformation 
$\gamma$=$15^\circ$.

The deformation of the $\beta$-$\gamma$ curve for SLy7 shows similar behavior 
except in the region of $\beta$=$0.15\mbox{-}
0.30$.  In the $\beta\sim0.30$ region, the deformation 
obtained for SLy7 is triaxial, which is different from the oblate deformation
with D1S. The wave functions in this region did not
affect the ND and SD bands, nor their side bands as shown later. 

Figure \ref{fig:density_beta} shows the density distribution of the
ND and SD minima, and confirms their large triaxial deformation. It also
reveals the deformed mean-field nature of the ND minimum and implies the
existence of a relationship between the SD minimum and the cluster structure. At the limit when all
${\bf Z}_i\rightarrow \vect{0}$ (all centroids of single particle wave
packets go to the origin of the coordinate frame), the AMD wave function is
identical to an eigenstate of the deformed harmonic oscillator whose
oscillation number is given by the relation 
$\omega_\sigma$=$2\hbar\nu_\sigma/M$. 
In contrast, when the system has
a cluster-like structure, the ${\bf Z}_i$ are separated into several groups to
describe cluster subunits.
In the case of the ND minimum, the centroids of the single particle wave
packets are located around the origin and the 
wave packets are strongly
deformed, $(\nu_x, \nu_y, \nu_z) = (0.16,0.14,0.11)$, suggesting
that the nature of the mean-field is deformed.  In the case of the SD minimum, 
the wave packets are further 
deformed, $(\nu_x, \nu_y, \nu_z) = (0.17,0.15,0.10)$, and reveal the aspect of a 
triaxially deformed mean-field.
In the case of the SD minimum, the centroids of the single particle wave packets 
appear separated into two (28+12) or
three ((12+16)+12) groups and the density distribution reveals an octupole
deformation implying that 
the SD minimum also has an asymmetric cluster-like nature. 
It will be shown using the $d$-constraint that this has a significant overlap with
the $^{12}{\rm C}$-$^{28}{\rm Si}$ cluster structure. 

There are few studies on the triaxial deformation of $^{40}{\rm Ca}$,
but some theoretical work has suggested the triaxiality 
of the ND \cite{ger69,ger77} and SD \cite{ina02} states.
The present calculation with the $\beta$-constraint has shown that most of the
states on the energy curve have a triaxial deformation. 
We therefore consider it indispensable to study the issue
without the assumption of spatial symmetry in order to understand the excited
states of $^{40}{\rm Ca}$.

\begin{figure}[hbt]
  \begin{tabular}{ccc}
    $\alpha$-$^{36}$Ar: type (a) & $\alpha$-$^{36}$Ar: type (b) &
   $^{12}$C-$^{28}$Si \\ 
    \begin{minipage}{0.15\textwidth}
      \begin{center}
	\includegraphics[width=\textwidth]{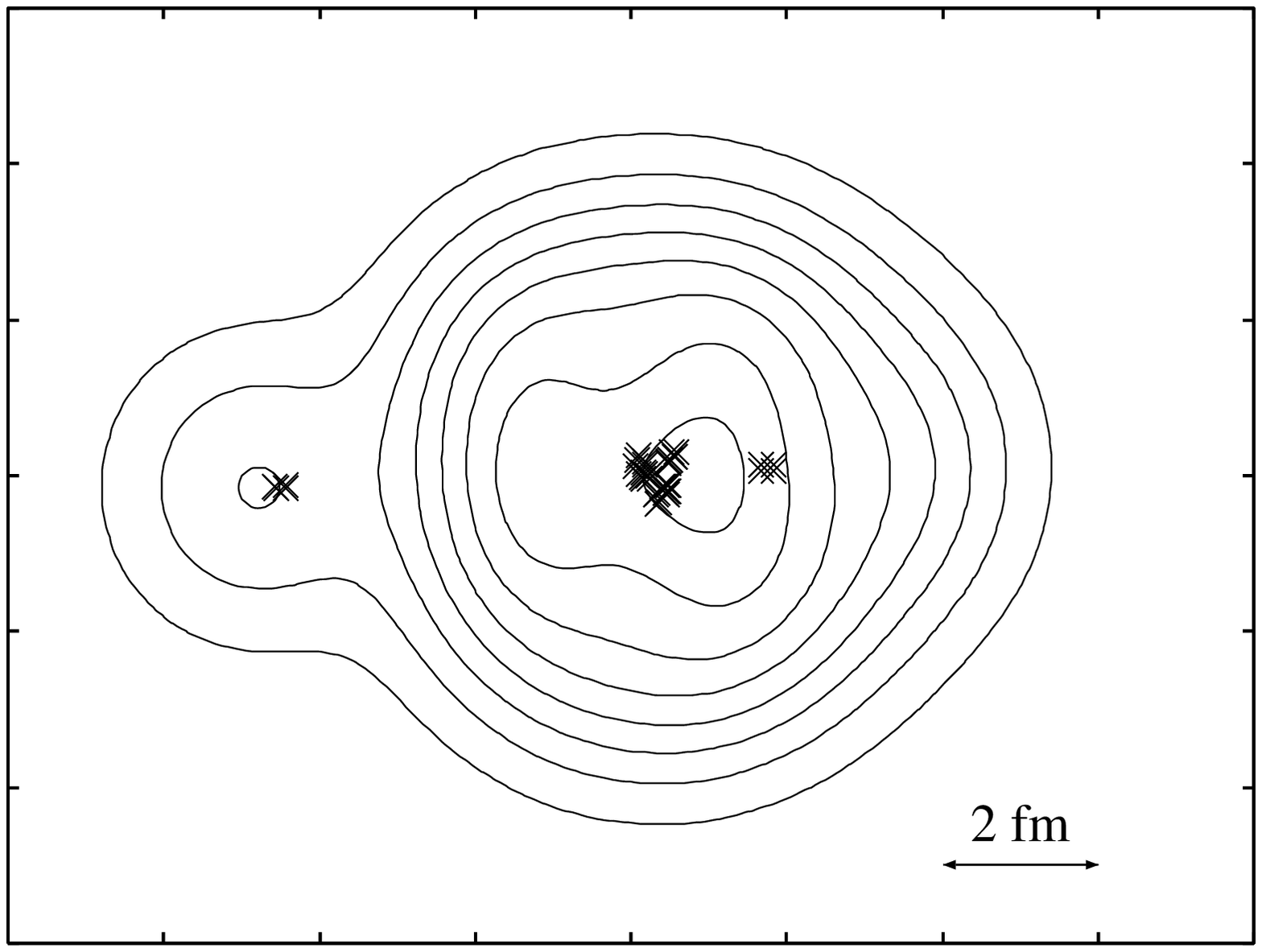}
      \end{center}
    \end{minipage}&
    \begin{minipage}{0.15\textwidth}
      \begin{center}
	\includegraphics[width=\textwidth]{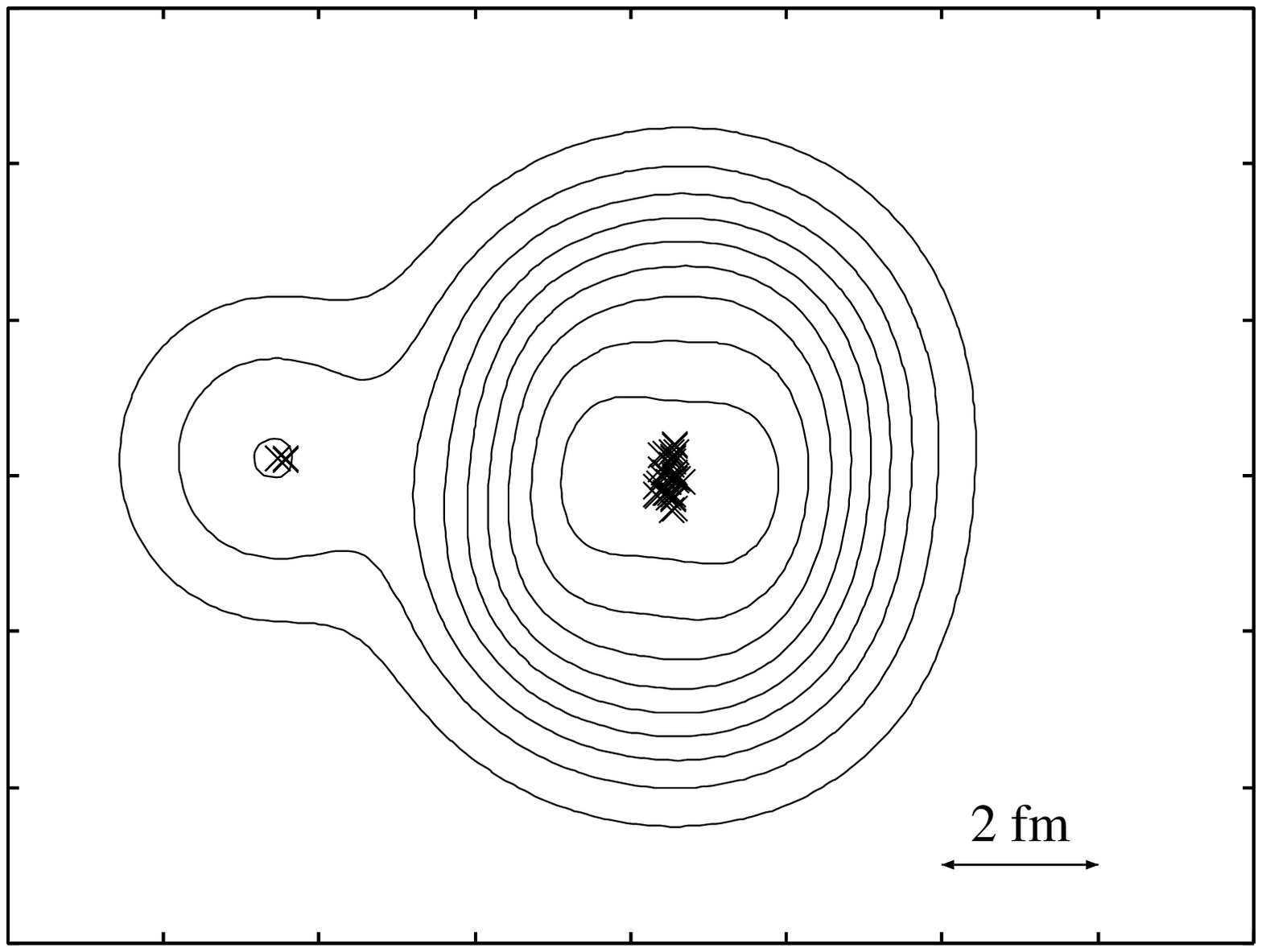}
      \end{center}
    \end{minipage}&
    \begin{minipage}{0.15\textwidth}
      \begin{center}
	\includegraphics[width=\textwidth]{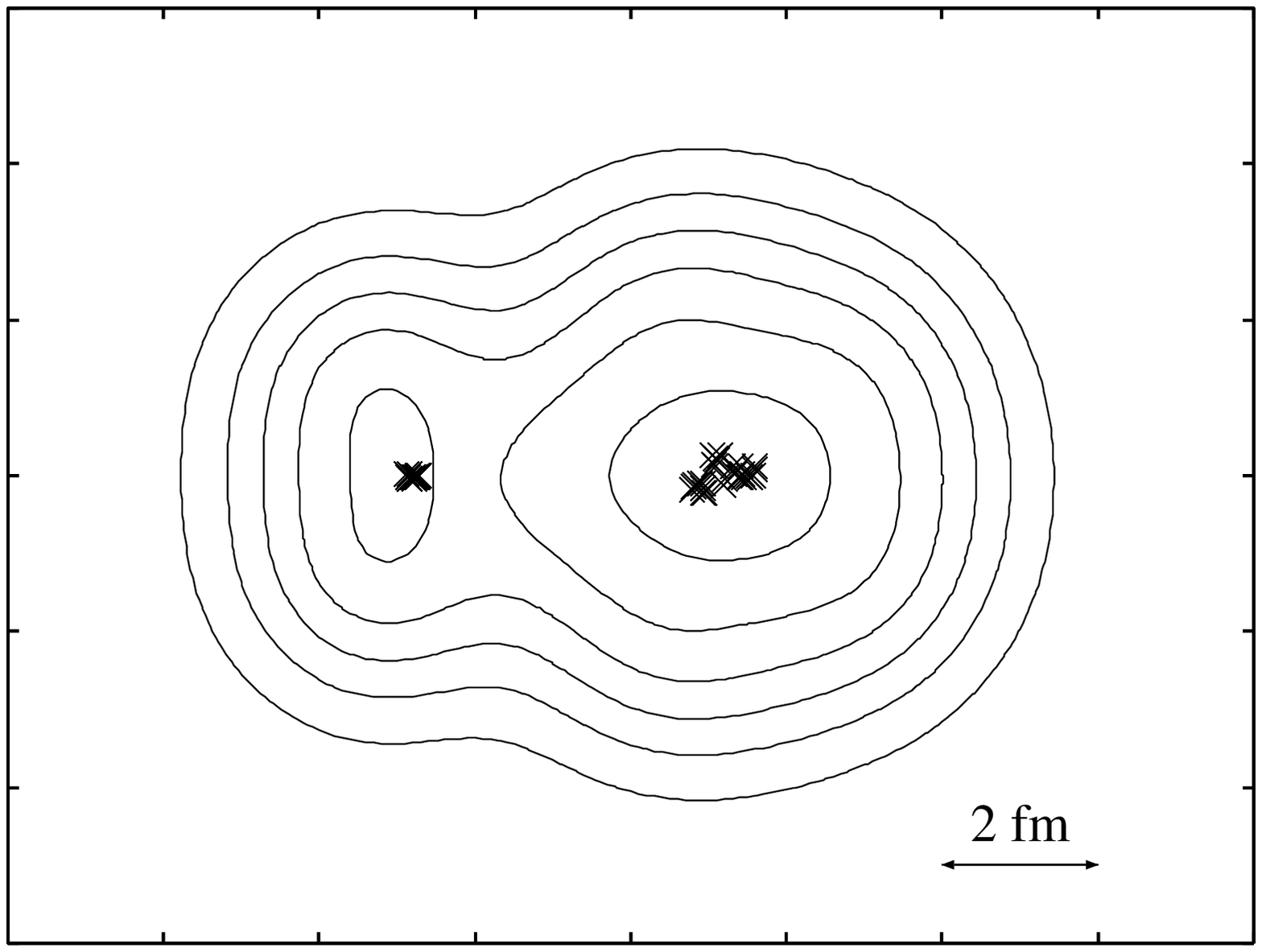}
      \end{center}
    \end{minipage}
  \end{tabular}
  \caption{    Density distributions of intrinsic states obtained with the
 $d$-constraint. $\mbox{\dAAr}$ is fixed to 5.0 fm for
 $\alpha$-$^{36}$Ar (a) and (b). $\mbox{\dCSi}$ is fixed to 4.0 fm for 
 $^{12}$C-$^{2}$Si. The crosses in the figure show the centroids of the
 wave packets. }  
  \label{fig:density_cluster}
\end{figure}

With the $\beta$-constraint, we do not find spatially developed 
clustering at the ND
and SD minima, although the density distribution of the SD state implies
its relationship to clustering. We have therefore applied
the $d$-constraint. We discuss the results obtained with \dAAr- and 
\dCSi-constraints. Other combinations of clusters such as
$2\alpha$-$^{32}$Si have been also applied, but they have appeared at comparatively 
high excitation energy levels and were not involved with the ND and SD states
even after the GCM calculation. 

By applying the \dAAr-constraint, we have obtained an excited energy curve
above the energy curve obtained with the $\beta$-constraint
(Fig. \ref{fig:surface}). Two different kinds of structure appeared on
the energy curve. In both cases, the system has well developed
$\alpha$-$^{36}$Ar cluster structures as may be clearly seen in their density
distributions (Fig. \ref{fig:density_cluster} (a) and (b)). The
difference between them is the orientation of the axis of symmetry in
the oblately deformed $^{36}$Ar cluster. The first is denoted as type
(a). In this type, the axis of symmetry for  $^{36}$Ar is perpendicular
to the vector that connects the $\alpha$ and $^{36}$Ar
clusters. Therefore the whole system has a triaxial deformation. This type
of structure is favored under \dAAr-constraints with large 
\dAAr values and appears in the region of
$\beta=$0.45-0.7. The second kind of structure is denoted type (b). In this type,
the axis of symmetry in the $^{36}$Ar cluster is parallel to the vector that
connects the $\alpha$ and $^{36}$Ar clusters, resulting in an axial
deformation of the system. This type is obtained when the inter-cluster
distance is restricted to a shorter distance (\dAAr=4.5-5.5 fm) and
appears in the region of $\beta$=0.15-0.3. It is interesting that
type (b) is bound deeper than type (a) for shorter inter-cluster 
distances and $^{36}$Ar changes its orientation as the inter-cluster
distance becomes larger. We assume that $^{36}$Ar changes its
orientation to make the overlap and potential energy between 
the $\alpha$ and $^{36}$Ar clusters larger.

By applying a \dCSi-constraint (\dCSi=4.0-6.0 fm), we obtained
a strongly deformed and excited energy curve that appears in the region of
$\beta$=0.55-0.8. The system has a prominent
$^{12}$C-$^{28}$Si cluster structure and is triaxially deformed as shown
in Fig. \ref{fig:density_cluster}. In this case, we did not find
a change in the orientation of the clusters. 

$d$-constraints have generated excited energy curves in which
the system has prominent cluster structures. The fact that these wave functions are mixed with the wave functions
obtained with the $\beta$-constraint and play an important role in describing
highly excited bands is discussed below. 

\subsection{Angular Momentum Projection}
\begin{figure}[hbt]
  \begin{center}
    \includegraphics[width=0.45\textwidth]{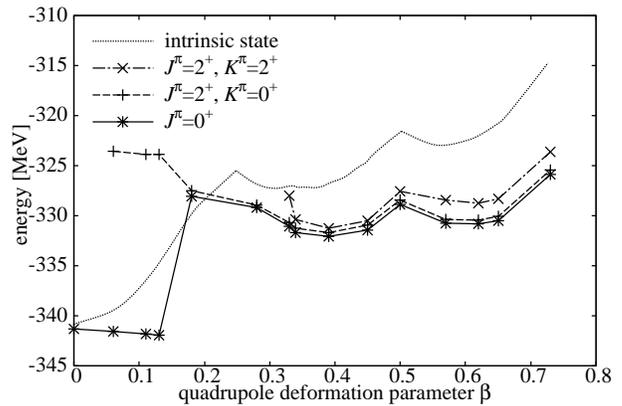}
    \caption{The energy curves of the $J^\pi = 0^+$ and $2^+$ states
   projected from wave functions obtained with the $\beta$-constraint with D1S.
   } \label{fig:AMP_MF} 
  \end{center}
\end{figure}

The wave functions obtained above are projected onto the eigenstate of
the total angular momentum. The $J^\pi$=$0^+$ and $2^+$ states obtained
with the $\beta$-constraint are shown in Fig. \ref{fig:AMP_MF}. AMP
reduces the excitation energies of the deformed states. For example,
the $J^\pi$=$0^+$ states of ND and SD are lowered by approximately 6 and 9 MeV
respectively. As a result, they almost degenerate in terms of energy. We have obtained two $2^+$
states for each given value of $\beta$ in the deformed region by the
diagonalization of the quantum number $K$. These states are
denoted as $K^\pi$=$0^+$ and $2^+$ according to the dominant component
of their wave function. The presence of the $K^\pi$=$2^+$ state is due
to the triaxial deformation in the $\beta \gtrsim 0.3$ region.
As will be discussed below,
the triaxial deformations of the ND and SD states leads to the presence of
their side bands, $K^\pi$=$2^+$.

The energy of the wave functions obtained with the $d$-constraint are also
lowered by AMP. The $\alpha$-$^{36}$Ar type (a), (b) and
$^{12}$C-$^{28}$Si states are lowered by approximately 
5-10, 10-15 MeV and 10 MeV respectively. $\alpha$-$^{36}$Ar type (a) and $^{12}$C-$^{28}$Si
wave functions have the $K^\pi$=$2^+$ components because of the triaxial
deformation.

\subsection{GCM Calculation}
\subsubsection{Energy levels and deformations}
After applying the AMP, we carried out GCM calculations. 
For the GCM basis, we adopted 15
$\beta$-constrained wave functions $\beta=0.00$-0.73, 7 
\dAAr-constrained wave functions of type (a) 
with $\mbox{\dAAr}=4.5$-9.0 fm, 3 \dAAr-constrained wave functions 
of type
(b) with $\mbox{\dAAr}=4.5$-5.5 fm, and
3 \dCSi-constrained wave functions with $\mbox{\dCSi}=4.0$-6.0 fm. Then we
obtained the final GCM wave functions by diagonalizing Hamiltonian and
norm matrices for the parity and angular momentum projected states with
$K^\pi=0^+$ and $\pm 2^+$ for the 28 independent wave functions.   

\begin{figure*}[hbt] 
  \begin{center}
    \includegraphics[width=0.90\textwidth]{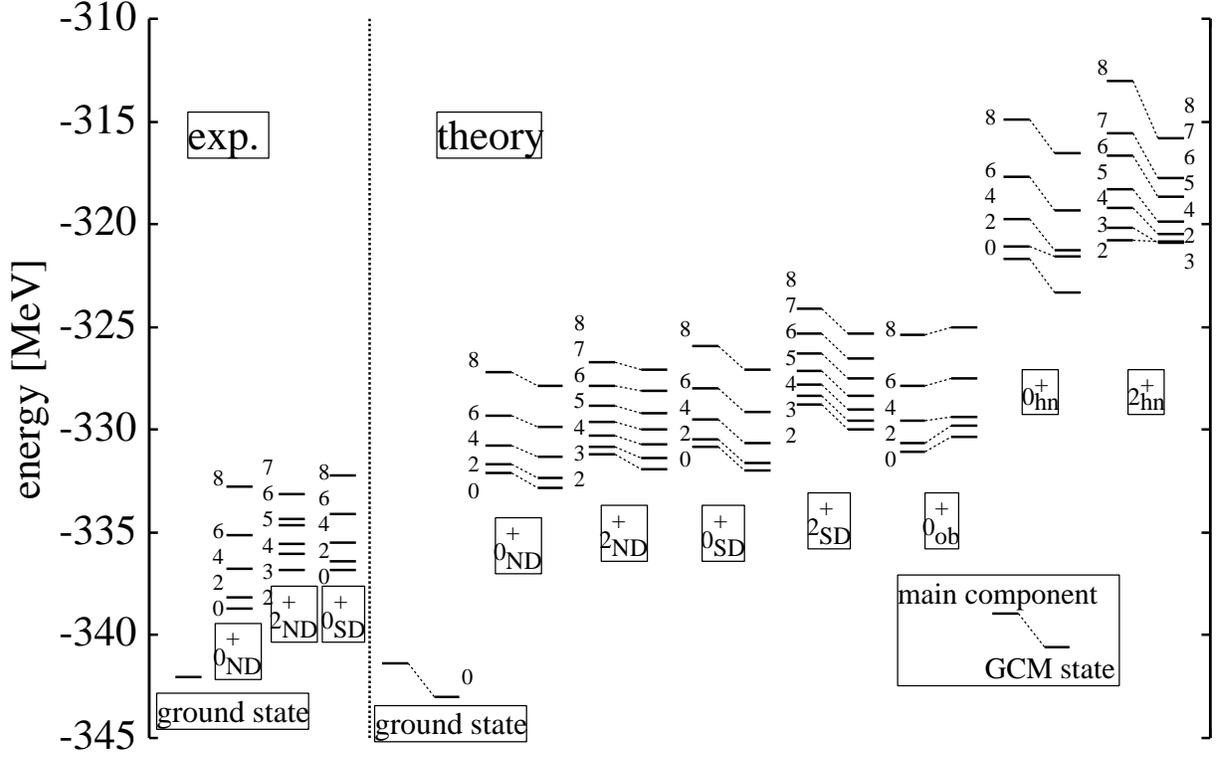}
    \caption{
      Energy levels in $^{40}$Ca. The left hand side represents experimental values, and the right hand side theoretical values. 
      The energies of superposed wave functions and that of the main component are represented. 
    }
    \label{fig:level}
  \end{center} 
\end{figure*}
The theoretical energy levels of the GCM states and the experimental
levels are shown in Fig. \ref{fig:level}.  The energies of the simple AMP
for the main components in the band heads are also given.
In most levels, the GCM states gain 1-2 MeV from the simple AMP.
 Experimentally, the $K^\pi = 0^+$, $2^+$ and $0^+$ bands built on
$J^\pi = 0_2^+$, $2_1^+$ and $0_3^+$ states are known to be the ND band
($K^\pi = 0_{\rm ND}^+$), the side band of the ND band ($K^\pi = 2_{\rm
ND}^+$) and the SD band ($K^\pi = 0_{\rm SD}^+$).  In the result of the
GCM calculation, we obtained three $K^\pi = 0^+$ and two $K^\pi = 2^+$
bands in low-lying states above the ground state.  We assigned the first
and second $K^\pi = 0^+$ bands to the observed $K^\pi = 0_{\rm ND}^+$
and $0_{\rm SD}^+$ bands respectively, and the first $K^\pi = 2^+$ band
to the observed $K^\pi = 2_{\rm ND}^+$ band, because the theoretical
moments of inertia and electric transition strength $B(E2)$ of these
bands correlate well with the experimental data for the corresponding bands as
discussed below.  The second $K^\pi = 2^+$ band in the results is
regarded as the side band $K^\pi = 2_{\rm SD}^+$ of the SD band.  We
denote the third theoretical $K^\pi = 0^+$ band as the $K^\pi = 0_{\rm ob}^+$
band because of the oblate shape.  We also obtained the $K^\pi = 0^+$
and $2^+$ bands in highly excited states with large $\alpha$-$^{36}$Ar
cluster structure components, which are the candidates for
$\alpha$-$^{36}$Ar higher-nodal bands, $K^\pi = 0_{\rm hn}^+$ and
$2_{\rm hn}^+$, observed experimentally \cite{yam94}.   


\begin{figure}[hbt]
  \begin{center}
    \includegraphics[width=0.45\textwidth]{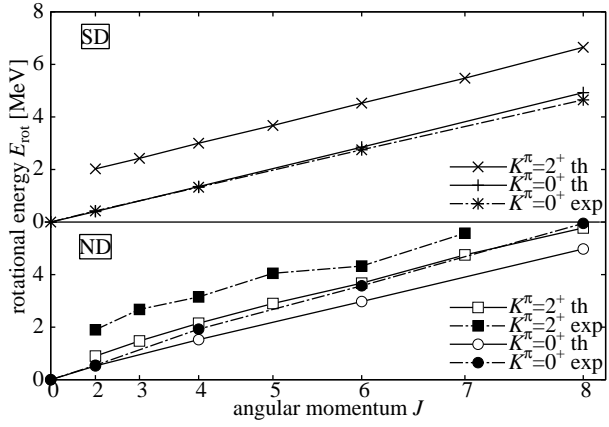}
    \caption{
      The rotational energies $E_{\rm rot}\equiv E-E_{0^+}$ of the ND and SD bands as functions of the angular momentum $J(J+1)$. 
    }
    \label{fig:moment}
  \end{center}
\end{figure}
Let us consider the GCM results of the low-lying states (ground, ND, SD and oblate states) in more detail by analyzing 
the squared overlap between $J^\pi = 0^+$ states 
and the AMP states from the $\beta$-constrained wave functions. 
The ND state's $K^\pi = 0_{\rm ND}^+$ and $2_{\rm ND}^+$ bands are mainly
constructed from the $\beta$-constrained wave functions around the ND
minimum occurring in the $\beta\sim 0.4$ region.
Their band-head states have a maximum overlap of approximately 85\% for the triaxially deformed state with $(\beta,\gamma)=(0.39,25.2^\circ)$. 
There is no mixing between the ND state and the ground state. 
The squared overlap with the $\beta$-constrained wave functions is almost unchanged up to high spin states along the $K^\pi = 0_{\rm ND}^+$ band. 
On the other hand, in the $K^\pi = 2_{\rm ND}^+$ band, the squared overlap of each component changes with the increase in spin, which implies the change of structure in this band. 
In the result of the GCM calculation, the theoretical excitation energies of the band heads in the $K^\pi = 0_{\rm ND}^+$ and $2_{\rm ND}^+$ bands are 11.2 MeV and 12.1 MeV respectively. 
These are much higher than the experimental excitation energies. 
However, the rotational energies in the band are reproduced well, as shown 
in Fig. \ref{fig:moment}. 

The SD states, $K^\pi = 0_{\rm SD}^+$ and $2_{\rm SD}^+$, are constructed mainly by $\beta$-constrained wave functions around the SD local minimum. 
The main component is the $\beta$-constrained wave function with triaxial shape,  $(\beta,\gamma)=(0.62,14.5^\circ)$. 
The squared overlap is more than 90\% in the band-head states, and almost unchanged up to high spin states along the $K^\pi = 0_{\rm SD}^+$ and $2_{\rm SD}^+$ bands. 
A small degree of mixing between the ND and SD bands is seen in the $\beta\sim 0.5$ region. 
The result of the GCM calculation reveals the band-head energies of the $K^\pi = 0_{\rm SD}^+$ and $2_{\rm SD}^+$ bands are 12.1 MeV and 14.0 MeV respectively. 
Although the present calculations overestimate
the experimental excitation energies of the $K^\pi = 0_{\rm SD}^+$ band
as well as the ND bands,
the rotational energies in the band are 
reproduced well as shown in Fig. \ref{fig:moment}. 

The oblate bands are dominated by the $\beta$-constrained wave functions in the $\beta\sim 0.3$ region. 
The main component is the oblate wave function with $(\beta,\gamma)=(0.33,51.2^\circ)$. 
The oblate band ($K^\pi = 0_{\rm ob}^+$) has not been discovered experimentally yet. 
Our prediction of the oblate band is consistent with spherical-basis AMD \cite{kan05} as well as SHF calculations \cite{ina02}. 

\begin{table}[bt]
  \begin{center}
    \caption{The squared overlap ($SO$) of $\beta$-constrained wave functions and $d$-constrained wave functions. } 
    \label{tab:components}
    \begin{tabular}{lccccc}
      \hline\hline
  &         &         & $\beta$-constraint & \multicolumn{2}{c}{$d$-constraint} \\
  \cline{5-6}
  & $K^\pi$ & $J^\pi$ & $SO$ & $SO$ & configuration \\
  \hline
  ND & $0_{\rm ND}^+$ & $0^+$   & 0.99 & 0.37 & $\alpha$-$^{36}$Ar  \\
  &         & $2^+$   & 0.99 & 0.40            \\
  &         & $4^+$   & 0.99 & 0.40             \\
  \cline{2-5}
  & $2_{\rm ND}^+$ & $2^+$   & 0.99 & 0.40             \\
  &         & $3^+$   & 0.99 & 0.45            \\
  &         & $4^+$   & 0.99 & 0.46            \\
  \hline
  SD & $0_{\rm SD}^+$ & $0^+$   & 0.95 & 0.59 & $^{12}$C-$^{28}$Si         \\
  &         & $2^+$   & 0.95 & 0.59             \\
  &         & $4^+$   & 0.95 & 0.56            \\
  \cline{2-5}
  & $2_{\rm SD}^+$ & $2^+$   & 0.95 & 0.61            \\
  &         & $3^+$   & 0.95 & 0.61            \\
  &         & $4^+$   & 0.95 & 0.60            \\
  \hline
  $\alpha$-$^{36}$Ar & $0_{\rm hn}^+$ & $0^+$ & 0.41 & 0.49 & $\alpha$-$^{36}$Ar\\
  higher-nodal &         & $2^+$   & 0.45 & 0.50 \\
  &         & $4^+$   & 0.52 & 0.50 \\
  \cline{2-5}
  & $2_{\rm hn}^+$ & $2^+$   & 0.38 & 0.55     \\
  &         & $3^+$   & 0.52 & 0.45          \\
  &         & $4^+$   & 0.59 & 0.39    \\
  \hline\hline
  \end{tabular}
\end{center}
\end{table}

\subsubsection{Cluster components}
In this section we discuss the contribution of the cluster 
wave functions in the ND and SD states. 
In Table \ref{tab:components}, we list the squared overlap ($SO$) values of the GCM states with the $\alpha$-$^{36}$Ar configuration space given by the set of \dAAr-constrained wave functions, and those with the $^{12}$C-$^{28}$Si configuration space, as well as the $SO$ values for the model space of the $\beta$-constrained wave functions. 
The definition of the specific functional space is explained in \S \ref{sec:SO}. 
The squared overlap between the ND state and $\beta$-constrained wave functions is 99\%. 
This means that the ND state is practically represented by the $\beta$-constrained wave functions alone. 
On the other hand, it is surprising that the ND states have a significant overlap of approximately 40\% with the \dAAr-constrained wave functions as well. 
This indicates that the $\beta$-constrained wave functions for the ND states include the $\alpha$-$^{36}$Ar component, though the spatially developed cluster structure is not seen in the density distributions. 
The $\alpha$-$^{36}$Ar cluster component in the ND band is mainly consistent with type (a) wave functions, while the squared overlap of type (b) 
wave functions in ND states is almost negligible. 
The result, that the ND state contains an $\alpha$-$^{36}$Ar cluster structure component, is associated with the results of the $\alpha$-$^{36}$Ar potential model calculation \cite{ohk88,rei90},$^{36}$Ar($^6$Li,$d$)$^{40}$Ca reaction \cite{yam93,yam94} and $\alpha$-$^{36}$Ar OCM calculation\cite{sak94}. 

The SD states are dominated by the $\beta$-constrained wave function as well as the ND states. 
However, the SD states also have a large overlap with the \dCSi-constrained wave functions as reflected by the $SO$ of approximately 60\%. 
We found that the $\beta$-constrained AMD wave functions for the SD states also  include the cluster components, even though the cluster structure is not visible in the density distributions. 
The \CSi\ cluster configurations also make a significant contribution to the energy of the SD states. 
In particular, the energies of the SD states gain 1-2 MeV due to the mixing of $^{12}$C-$^{28}$Si cluster structure wave functions. 
This is associated with the results of spherical-basis AMD \cite{kan05} according to which the SD states have a $^{12}$C-$^{28}$Si cluster structure configuration. 

\subsubsection{$\alpha$-$^{36}$Ar higher-nodal states, $K^\pi = 0_{\rm hn}^+$ and $2_{\rm hn}^+$ bands}
As shown in Fig. \ref{fig:level}, 
we obtained $K^\pi = 0^+$ and $2^+$ bands 
with large $\alpha$-$^{36}$Ar cluster components 
in the excitation energy region approximately 10 MeV higher than the ND band.
The main component of the higher $\alpha$-$^{36}$Ar bands, $K^\pi = 0_{\rm hn}^+$ and $2_{\rm hn}^+$, are \dAAr-constrained wave functions obtained with a large distance $\mbox{\dAAr}=6.0$ fm. 
We assume that this corresponds to the $\alpha$-$^{36}$Ar higher-nodal band observed with the $^{36}$Ar($^6$Li,$d$)$^{40}$Ca reaction \cite{yam94}, 
where the fragments of the $J^\pi = 0^+$ state for this band
were reported
around 8 MeV above the band-head of the ND band.

\begin{figure}[hbt]
  \begin{center}
    \includegraphics[width=0.45\textwidth]{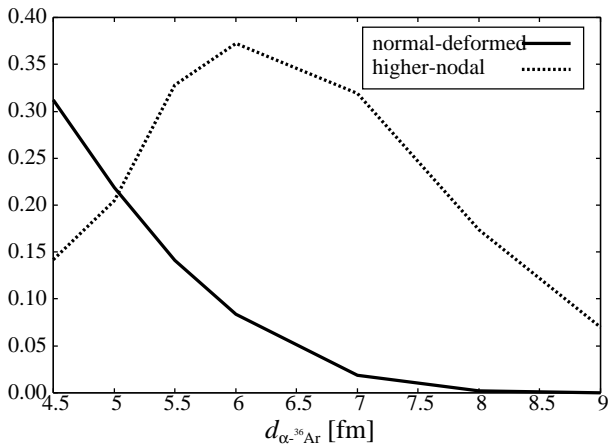}
    \caption{
      The squared overlap of the $J^\pi = 0^+$ states in the ND band ($K^\pi = 0_{\rm ND}^+$) and the $\alpha$-$^{36}$Ar higher-nodal band ($K^\pi = 0_{\rm hn}^+$) with \dAAr-constrained wave functions. } 
    \label{fig:A36Ar_ovlp}
  \end{center}
\end{figure}
The squared overlap between \dAAr-constrained wave functions and the $J^\pi = 0^+$ states in $K^\pi = 0_{\rm ND}^+$ and $0_{\rm hn}^+$ bands are shown in Fig. \ref{fig:A36Ar_ovlp}. 
The squared overlap for $0_{\rm hn}^+$ 
is suppressed in the small \dAAr\ region and 
has a peak at $\mbox{\dAAr}=6$ fm, which demonstrates the nodal 
property of the inter-cluster motion in this band.
It can be assumed that the $\alpha$-$^{36}$Ar higher-nodal 
states arise from the inter-cluster excitation built on the ND states. 
In other words, the significant component of the $\alpha$-$^{36}$Ar 
cluster structure in the ND states must be essential 
for the formation of the higher-nodal states, 
because the cluster component in the small distance region is possible if the ND states contain no cluster component. 
This means that the higher-nodal states appear as a consequence of the orthogonality to the cluster components with a small distance $d$, which are already contained in the ND states. 
This situation is similar to the relationship between the ground band and $\alpha$-$^{40}$Ca higher-nodal band in $^{44}$Ti suggested by Kimura {\it et al.} with deformed-basis AMD \cite{kim06}. 

\subsubsection{Electric transitions}
\begin{figure*}
  \begin{center}
    \includegraphics[width=\textwidth]{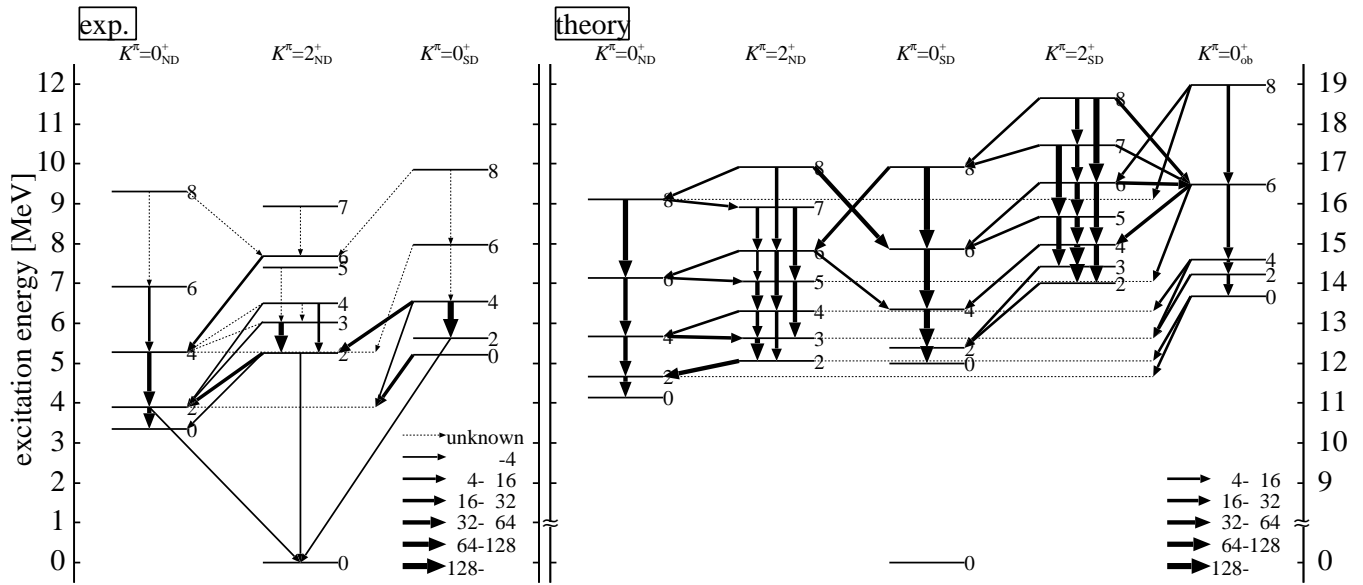}
    \caption{
      Experimental and theoretical $B(E2)$ values. 
      Units are Weisskopf units, $B(E2)_{\rm W. u.} = 8.12~e^2{\rm fm}^4$ for $^{40}$Ca. 
      In theoretical values, transitions stronger than 4 4 $B(E2)_{\rm W.\ u.}$ are presented. 
      The experimental data are taken from Refs.~\onlinecite{sak94} and \onlinecite{NDS}. 
      ``unknown'' indicates the transition has been observed but $B(E2)$ has not been measured.}
    \label{fig:level_gamma}
  \end{center}
\end{figure*}
Here, we investigate the electric quadrupole strengths and discuss the band structure. 
The theoretical and experimental values of $E2$ transition strengths $B(E2)$ are shown in Fig. \ref{fig:level_gamma}. 
The $B(E2)$ values for intra-band transitions are remarkably strong, and the transitions between the $K^\pi = 0_{\rm ND}^+$ and $K^\pi = 2_{\rm ND}^+$ bands, and those between the $K^\pi = 0_{\rm SD}^+$ and $2_{\rm SD}^+$ bands are also strong. 
This reflects the side-band features of the 
$K^\pi = 2_{\rm ND}^+$ and $K^\pi = 2_{\rm SD}^+$ bands, corresponding to $K^\pi = 0_{\rm ND}^+$ and $K^\pi = 0_{\rm SD}^+$ respectively. 
The inter-band transitions between ND and SD states
and those from ND or SD states to the ground state are underestimated. 
This is because mixing of wave functions among these bands 
is small in our calculation. 
On the other hand, we predict the transitions 
from the oblate $K^\pi = 0_{\rm ob}^+$ band to the ND bands, 
$K^\pi = 0_{\rm ND}^+$ and $2_{\rm ND}^+$, because of the
mixing among these bands. 

\begin{table}
  \begin{center}
    \caption{
      Theoretical and experimental $B(E2)$ of $^{40}$Ca. 
      The units of transitions are Weisskopf units, $B(E2)_{\rm W. u.} = 8.12\  e^2 {\rm fm}^4$. 
      The asterisk-marked experimental $B(E2)$ values are taken from Ref.~\onlinecite{sak94}, and other values are taken from Ref.~\onlinecite{NDS}. 
    }
\label{tab:BE2}
\begin{tabular}{ccccc}
  \hline\hline
$K^\pi_i \rightarrow K^\pi_f$  & $I_i$ & $I_f$ & $B(E2)_{\rm th}$ & $B(E2)_{\rm exp}$ \\
  \hline
    $0_{\rm ND}^+ \rightarrow \mbox{g. s. }$ & $2^+$ & $0^+$ & $<0.05$ & $2.26\pm 0.14$\\
  \hline
    $2_{\rm ND}^+ \rightarrow \mbox{g. s. }$ & $2^+$ & $0^+$ &$<0.05$ & $0.13\pm 0.04$ \\
  \hline
  $0_{\rm SD}^+ \rightarrow \mbox{g. s. }$ & $2^+$ & $0^+$ &  $<0.05$ & $0.20\pm 0.05$\\
  \hline
  $0_{\rm ND}^+ \rightarrow 0_{\rm ND}^+$ & $2^+$ & $0^+$ &  39.0 & $32\pm 4$\\
  & $4^+$ & $2^+$ & 55.5 & $61\pm 10$\\
  & $6^+$ & $4^+$ & 63.7 & $17^{+9}_{-17}$\\
  & $8^+$ & $6^+$ & 65.1 & ---\\
  \hline
  $0_{\rm SD}^+ \rightarrow 0_{\rm SD}^+$ & $2^+$ & $0^+$ & 118.5 & ---\\
  & $4^+$ & $2^+$ & 166.0 & $170\pm 40$\\
  & $6^+$ & $4^+$ & 178.8 & ---\\
  & $8^+$ & $6^+$ & 145.2 & ---\\
  \hline
  $2_{\rm ND}^+ \rightarrow 2_{\rm ND}^+$ & $3^+$ & $2^+$ & 66.6 & $>71$, $82\pm 26^*$\\
  & $4^+$ & $2^+$ & 22.0 & $23\pm 5^*$ \\
  & $4^+$ & $3^+$ & 23.6 & $< 1300^*$ \\
  \hline
  $2_{\rm ND}^+ \rightarrow 0_{\rm ND}^+$ & $2^+$ & $0^+$ & $<0.05$ & $1.3\pm 0.4$, $0.54\pm 0.14^*$\\
  & $2^+$ & $2^+$ & 44.0 & $22\pm 6$, $25\pm 6^*$\\
  & $3^+$ & $2^+$ & 0.1 & $3.7\pm 0.7^*$ \\
  & $3^+$ & $4^+$ & 23.3 & $<22^*$\\
  & $4^+$ & $2^+$ & 0.3 & $3.8\pm 0.8$\\
  & $4^+$ & $4^+$ & 14.7 & $6.8\pm 5.2^*$\\
  & $6^+$ & $4^+$ & $<0.05$ & $4.2\pm 1.0$\\
  \hline
  $0_{\rm SD}^+ \rightarrow 0_{\rm ND}^+$ & $0^+$ & $2^+$ & 0.1 & $17\pm 3$\\
  & $2^+$ & $0^+$ & $<0.05$ & $2.6\pm 0.7$\\
  & $4^+$ & $2^+$ & $<0.05$ & $2.6\pm 0.6$\\
  \hline
  $0_{\rm SD}^+ \rightarrow 2_{\rm ND}^+$ & $4^+$ & $2^+$ & 0.2 & $22\pm 6$\\
  \hline\hline
\end{tabular}
\end{center}
\end{table}
Detailed comparisons between the theoretical values and experimental data of the $B(E2)$ values are shown in Table \ref{tab:BE2}. 
The $B(E2)$ values for the intra-band transitions in the ND and SD states are reproduced well, except for the $6^+\rightarrow 4^+$ transition when $K^\pi = 0_{\rm ND}^+$. 
The small value of the experimental $B(E2; 6^+\rightarrow 4^+)$ is overestimated by the calculation, although the error in the experimental value is rather large. 

\parag{Quadrupole moment}
\begin{table*}
  \begin{center}
    \caption{
      Intrinsic quadrupole moments of the ND band and the SD band extracted from $B(E2)$. 
      For theoretical study, (1), (2), (3) and (4) are taken from the results of spherical-basis AMD \cite{kan05}, HFB-BCS + GCM \cite{ben03}, shell model \cite{cau07} and $\alpha$-$^{36}$Ar OCM \cite{sak94}, respectively. 
      For experimental data, (I) is obtained from life-time and branching ratio, asterisk-marked values are taken from Ref.~\onlinecite{sak94} and others are from Ref.~\onlinecite{NDS}. 
      (II)-a and (II)-b are taken from Refs.~\onlinecite{ide01} and \onlinecite{chi03} respectively, which are obtained through a global fitting of all the transitions. 
    }
\label{tab:comparison}
\begin{tabular}{cccccccccccc}
\hline\hline
\multicolumn{2}{c}{$K^\pi$}   && \multicolumn{5}{c}{theory} && \multicolumn{3}{c}{experiment}\\
\cline{1-2}\cline{4-8}\cline{10-12}
$I_i$ & $I_f$ && present work & (1) & (2) & (3) & (4) && (I) & (II)-a & (II)-b \\
\hline
\multicolumn{2}{c}{$0_{\rm ND}^+$} &&& A3 & & & &&&&  band 2   \\
\cline{1-2}
$2^+$ & $0^+$ && 126.2 & 57 & 75.2 & 121 & 117 && $110\pm 10$& & $74\pm 14$ \\
$4^+$ & $2^+$ && 125.9 & 64 & 23.9 & 118 & 117 && $130\pm 10$& & $74\pm 14$ \\
$6^+$ & $4^+$ && 128.5 & 64 & 77.4 & 105 & 117 && ~$66^{+16}_{-66}$ & & $74\pm 14$ \\
$8^+$ & $6^+$ && 127.0 & 78 &      &  83 & 115 &&  & &$74\pm 14$  \\
\hline
\multicolumn{2}{c}{$2_{\rm ND}^+$}&& & & & & & & &&  \\
\cline{1-2}
$3^+$ & $2^+$ && 123.4 &&&& 119 && $140\pm 20^\ast$\\
$4^+$ & $2^+$ && 122.7 &&& 106 & 115 && $120\pm 10^\ast$\\
$4^+$ & $3^+$ && ~85.1  &&&& 115 && $< 630^\ast$\\
\hline
\multicolumn{2}{c}{$0_{\rm SD}^+$}&&& A3 (A2) & & & &&&  band 1 & band 1  \\
\cline{1-2}
$2^+$ & $0^+$ && 219.9 &  129 & 133.9 & 171 & &&& $180^{+39}_{-29}$ & $130\pm 5$ \\
$4^+$ & $2^+$ && 217.8 &  129 & 97.6  & 169 & &&$220\pm 26$& $180^{+39}_{-29}$ & $130\pm 5$  \\
$6^+$ & $4^+$ && 215.3 &  121 & 160.2 & 167 & &&& $180^{+39}_{-29}$ & $130\pm 5$\\
$8^+$ & $6^+$ && 189.7 &  97 (70)& 157.9 & 166 & &&& $180^{+39}_{-29}$ & $130\pm 5$\\
\hline\hline
\end{tabular}
\end{center}
\end{table*}
The theoretical and experimental values of quadrupole moments deduced from the $E2$ strengths for the $K^\pi = 0_{\rm ND}^+,\  2_{\rm ND}^+$ and $0_{\rm SD}^+$ bands are listed in Table \ref{tab:comparison}. 
The results of other theoretical studies are also shown in the table. 
The intrinsic quadrupole moment $Q$ is defined by the $B(E2)$ value as follows. 
\begin{equation}
Q \equiv \sqrt{\frac{16\pi}{5e^2}\frac{B(E2,\  I_i\rightarrow I_f)}{\ovlp{I_iK20}{I_fK}^2}}, 
\label{eq:Q}
\end{equation}
where $\ovlp{I_iK20}{I_fK}$ is the Clebsh-Gordan coefficient. 

The theoretical quadrupole moments of the ND and SD states are around 130 fm$^2$ and 220 fm$^2$ respectively. 
The values are almost unchanged up to high spin. 
This means that the ND and SD bands consist of approximately rigid rotor states. 
These results are consistent with the rigid rotor-like spin dependence of the excitation energies of the ND and SD states shown in Fig. \ref{fig:moment}.
%
The theoretical $Q$ moments are consistent with experimental values evaluated from life-time measurements and branching ratios (I) except for the $6^+ \rightarrow 4^+$ transition in the ND state. 
The small value of the experimental $Q$ moment for the $6^+\rightarrow 4^+$ transition might suggest some structure change in high spin states in the ND band. 

We compared our results with those of other theoretical studies. 
The $Q$ values in (4) $\alpha$-$^{36}$Ar OCM are consistent with the 
present values for the ND states. 
This may indicate that the deformation of the ND states in our calculations
is similar to that calculated with the $\alpha$-$^{36}$Ar cluster model. 
In (1) the spherical-basis AMD and (3) the shell model, smaller $Q$ values for the $K^\pi = 0^+$ states were obtained by comparison to our model. 
In (2) SHF-BCS + GCM, the $Q$ values are also smaller than our results for $K^\pi = 0^+$ states. 
In particular, the $Q$ moments for the $4^+ \rightarrow 2^+$ transition in both the ND and SD states are remarkably small.
This shows that the ND and SD bands obtained in the SHF-BCS + GCM calculation are not rigid rotor-like. 
This seems to be inconsistent with experimental results regarding the rigid rotor-like property, which has been found in rotational energies and $E2$ transitions. 
In recent experimental observations, the $Q$ values have been obtained by globally fitting all transitions. 
They ((II)-a and b) are smaller than the values of the life-time measurement (I) and also smaller than the present results. 
Further analysis requires more detailed measurements of the $E2$ transition strengths. 

\parag{Triaxial rotor}

In this section we discuss the characteristics of the side bands, $K^\pi = 2^+$, in the ND and SD states. 
In a naive collective model, a $K^\pi = 2^+$ band can be described by either a triaxial rotor or a $\gamma$ vibration. 
In order to identify the mode of the side bands, it is useful to analyze the ratio of $E2$ strength for $K^\pi = 2^+$ to $K^\pi = 0^+$ transitions, 
\begin{equation}
R=
\frac{
B(E2; (2^+, K^\pi = 2^+)\rightarrow (0^+, K^\pi = 0^+))
}
{
B(E2; (2^+, K^\pi = 2^+)\rightarrow (2^+, K^\pi = 0^+))
}
,  \label{R}
\end{equation}
where $B(E2; (J_i^\pi, K_i^\pi)\rightarrow (J_f^\pi, K_f^\pi))$ is the value of quadrupole electric transition strength from a $J_i^\pi$ state in the $K_i^\pi$ band to a $J_f^\pi$ state in the $K_f^\pi$ band. 
The $R$ value is set to the constant value $7/10$ in the $\gamma$ vibration mode, while it is a function of the static $\gamma$ parameter in the triaxial rotor mode \cite{dav58},
\begin{equation}
  R^{\rm rot} (\gamma) = 
\frac{7}{20} 
\left.
\left[ 1 - \frac{3-2\sin^2(3\gamma)}{\sqrt{9-8\sin^2 (3\gamma)}} \right]
\right/
\frac{\sin^2 (3\gamma)}{9-8\sin^2 (3\gamma)}
. 
\end{equation}

\begin{figure}[hbt]
  \includegraphics[width=0.45\textwidth]{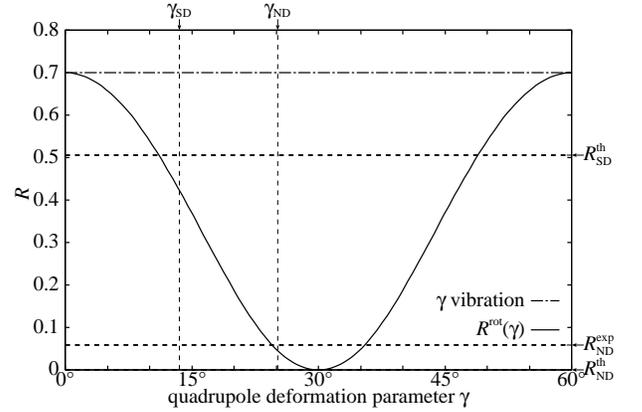}
  \caption{
 The values of the ratio of $E2$ strength $R$ 
as a function of $\gamma$ value. 
The dot-dashed line and the solid line show the limiting values 
of the $\gamma$-vibration and the triaxial rotor respectively. 
The $R_{\rm ND}^{\rm th} (\sim 0)$ and $R_{\rm SD}^{\rm th} (= 0.51)$ values are the theoretical $R$ values for the ND and SD states derived from the $E2$ transition strengths respectively, 
while $R_{\rm ND}^{\rm exp}$ is that of the experiment for the ND state. 
The deformation parameters 
$\gamma_{\rm ND}(=25.2^\circ)$ and $\gamma_{\rm SD}(=14.5^\circ)$ 
for the main components of the ND and SD states are also
shown.
}
  \label{fig:davydov}
\end{figure}
The function $R^{\rm rot} (\gamma)$ is plotted as a function of $\gamma$ value in Fig. \ref{fig:davydov}. 
The value of $R^{\rm rot}$ is 0.7 for the axial symmetric limit, $\gamma\rightarrow 0^\circ$ and $60^\circ$. 
As the triaxiality increases, $R^{\rm rot}$ decreases, going to zero at the peak of triaxiality, $\gamma=30^\circ$. 
Regarding the ND state, the theoretical value $R$ of the $E2$ strength ratio is nearly equal to 0, and the experimental value $R=0.06$ is also close to 0.
Both the values agree well with 
$R^{\rm rot} (\gamma\sim 30^\circ)$ 
but are far from that of the $\gamma$ vibration limit, 0.7. 
In the present results, the $\gamma$ value of the main component 
of the ND state is $25.2^\circ$, which is consistent with 
$\gamma\sim 30^\circ$ for the triaxial rotor
derived from the experimental $R$ value. 
We therefore conclude that the mode of the observed 
$K^\pi = 2_{\rm ND}^+$ band 
is a triaxial rotor, not a $\gamma$ vibration. 

Regarding the SD state, the theoretical value of $R$ is equal to 0.51, and the $\gamma$ value of the main component of the theoretical SD state is $14.5^\circ$. These values,
$(\gamma,R) = (14.5^\circ,0.51)$ can be described 
by the relation $R^{\rm rot}(\gamma)$ in Eq. (\ref{R}) for the triaxial rotor.
This indicates that the case of $K^\pi = 2_{\rm SD}^+$ in the SD state 
is the side band of the triaxial rotor. 
The $K^\pi = 2_{\rm SD}^+$ band has not yet been observed experimentally. 
In order to confirm the triaxiality of the SD band, observations of the $K^\pi = 2_{\rm SD}^+$ band and the $E2$ strength for the inter-band transitions are required. 

\section{Summary}
\label{sec:summary}
We investigated the ground state and excited states of $^{40}$Ca in the framework of deformed-basis AMD focusing on the aspects of triaxiality and clustering in the deformed states. 
Superposing mean-field-type and cluster structure-type wave functions obtained with $\beta$- and $d$-constraints respectively, we obtained the ground states, ND, SD, oblate and $\alpha$-$^{36}$Ar higher-nodal states. 
We found both the ND and SD bands are constructed from 
triaxially deformed shapes, and we obtained the $K^\pi = 2^+$ 
side bands of these bands. 
The theoretical $B(E2)$ and moments of inertia are consistent with 
experimental data, although the excitation energies 
are higher than empirical values. 
By analyzing quadrupole moments calculated from the $B(E2)$ and the $E2$ strength ratio $R$ values, we found that the ND and SD states can be understood by a relatively rigid rotor conception of triaxial deformations. 
The ND band and it's side band contain approximately 40\% of the
$\alpha$-$^{36}$Ar cluster structure component, and the SD band and it's
side band contain approximately 60\% of the $^{12}$C-$^{28}$Si cluster structure
component.
The $\alpha$-$^{36}$Ar higher-nodal band was obtained due to the
excitation of inter-cluster motion between the $\alpha$ and $^{36}$Ar
clusters. 
The present results suggested that cluster correlation will be 
important for deformation and excitation
even in medium- and heavy-weight nuclei. 

\section*{Acknowledgments}
This work has been supported by JSPS Research Fellowships for Young Scientists. 
The numerical calculations were carried out on SX8 at YITP in Kyoto University and on SX5 and SX8 at RCNP in Osaka University. 
The authors would like to thank Dr. Inakura and Dr. Fujiwara for fruitful discussions. 
This work is supported by a Grant-in-Aid for the 21st Century COE ``Center for Diversity and Universality in Physics'' from the Ministry of Education, Culture, Sports, Science and Technology of Japan (MEXT).


\begin{thebibliography}{99}
\bibitem{hor72} H. Horiuchi, K. Ikeda and Y. Suzuki, Prog. Theor. Phys. Suppl. No. 52, 89 (1972), and references therein. 
\bibitem{fuj80} Y. Fujiwara, H. Horiuchi, K. Ikeda, M. Kamimura, K. Kat$\bar{\rm o}$, Y. Suzuki and E. Uegaki, Prog. Theor. Phys. Suppl. No. 68, 29 (1980), and references therein. 
\bibitem{mic98} F. Michel, S. Ohkubo and G. Reidemeister, Prog. Theor. Phys. Suppl. No. 132, 7 (1998), and references therein. 
\bibitem{sak98} T. Sakuda and S. Ohkubo, Prog. Theor. Phys. Suppl. No. 132, 103 (1998), and references therein. 
\bibitem{yam98} T. Yamaya, K. Katori, M. Fujiwara, S. Kato and S. Ohkubo, Prog. Theor. Phys. Supp., No. 132, 73 (1998), and references therein. 
\bibitem{ger67} W. J. Gerace and A. M. Green, Nucl. Phys. {\bf 93}, 110 (1967). 
\bibitem{ger69} W. J. Gerace and A. M. Green, Nucl. Phys. {\bf A123}, 241 (1969). 
\bibitem{ger77} W. J. Gerace and J. P. Mestre, Nucl. Phys. {\bf A285}, 253 (1977). 
\bibitem{pal80} K. F. Pal and R. G. Lovas, Phys. Lett. {\bf B96}, 19 (1980). 
\bibitem{ohk88} S. Ohkubo and K. Umehara, Prog. Theor. Phys. {\bf 80}, 598 (1988). 
\bibitem{rei90} G. Reidemeister, S. Ohkubo and F. Michel, Phys. Rev. {\rm C41}, 63 (1990). 
\bibitem{oga77} T. Ogawa, Y. Suzuki and K. Ikeda, prog. Theor. Phys. {\bf 57}, 1072 (1977). 
\bibitem{sak94} T. Sakuda and S. Ohkubo, Phys. Rev. {\bf C49}, 149 (1994). 
\bibitem{yam93} T. Yamaya, M. Saito, M. Fujiwara, T. Itahashi, K. Katori, T. Suehiro, S. Kato, S. Hatori and S. Ohkubo, Phys. Lett. {\bf B306}, 1 (1993). 
\bibitem{yam94} T. Yamaya, M. Saitoh, M. Fujiwara, T. Itahashi, K. Katori, T. Suehiro, S. Kato, S. Hatori and S. Ohkubo, Nucl. Phys. {\bf A573}, 154 (1994). 
\bibitem{mid72} R. Middleton, J. D. Garrett and H. T. Fortune, Phys. Lett. {\bf B39}, 339 (1972). 
\bibitem{mac71} J. R. MacDonald, D. H. Wilkinson and  D. E. Alburger, Phys. Rev. {\bf C3}, 219 (1971), and references therein.  
\bibitem{ide01} E. Ideguchi {\it et al.}, Phys. Rev. Lett. {\bf 87}, 222501 (2001). 
\bibitem{ina02} T. Inakura, S. Mizutori, M. Yamagami and K. Matsuyanagi, Nucl. Phys. {\bf A710}, 261 (2002). 
\bibitem{ben03} M. Bender, H. Flocard and P. -H. Heenen. Phys. Rev. {\bf C68}, 044321 (2003). 
\bibitem{kan05} Y. Kanada-En'yo and M. Kimura, Phys. Rev. {\bf C72}, 064322 (2005). 
\bibitem{cau07} E. Caurier, J. Menendez, F. Nowacki, A. Poves, Phys. Rev. {\bf C75}, 054317 (2007). 
\bibitem{tan04} Y. Taniguchi, M. Kimura and H. Horiuchi, Prog. Theor. Phys. {\bf 112}, 475 (2004). 
\bibitem{dot97} A. Dote, H. Horiuchi and Y. Kanada-En'yo, Phys. Rev. {\bf C56}, 1844 (1997). 
\bibitem{kan95} Y. Kanada-Enyo and H. Horiuchi, Prog. Theor. Phys. {\bf 93}, 115 (1995). 
\bibitem{NDS} J. A. Cameron and B. Singh, Nuclear Data Sheets 102, 293 (2004). 
\bibitem{chi03} C. Chiara {\it et al.},  Phys. Rev. {\bf C67}, 041303 (2003). 
\bibitem{kim06} M. Kimura, H. Horiuchi, Nucl. Phys. A767, 58 (2006). 
\bibitem{dav58} A. S. Davydov and G. F. Filippov, Nucl. Phys. {\bf 8}, 237 (1958). 
\end{thebibliography}
\end{document}